\documentclass[showpacs,aps,twocolumn]{revtex4}
\usepackage{epsfig}
\usepackage{graphicx}
\usepackage{amsmath,amssymb,amsfonts}
\usepackage{array}
\usepackage{url}
\usepackage{hyperref}
\usepackage{multirow}
\usepackage{adjustbox}

\usepackage{lineno}
\usepackage{xspace}
\usepackage[usenames,dvipsnames]{color}  

\newcommand{\bef}{\begin{figure}}
\newcommand{\eef}{\end{figure}}
\newcommand{\bc}{\begin{center}}
\newcommand{\ec}{\end{center}}

\newcommand{\be}{\begin{equation}}
\newcommand{\ee}{\end{equation}}
\newcommand{\bea}{\begin{eqnarray}}
\newcommand{\eea}{\end{eqnarray}}

\begin{document}
%\title{Multiplicity Dependence of heat capacity, Conformal Symmetry Breaking and speed of sound in small system}
\title{Study of QCD dynamics using small systems}
%\title{Systematics approach in understanding role of randomization in small system}
\author{Suman Deb}
\email{sumandeb0101@gmail.com}
\author{Golam Sarwar}
\email{golamsarwar1990@gmail.com}
\author{Raghunath Sahoo}
\email{Raghunath.Sahoo@cern.ch (Corresponding author, Presently CERN Scientific Associate, CERN, Geneva, Switzerland)}
\affiliation{Department of Physics, Indian Institute of Technology Indore, Simrol, Indore 453552, INDIA}
\author{Jan-e Alam}
\email{jane@vecc.gov.in}
\affiliation{Variable Energy Cyclotron Centre, 1/AF, Bidhan Nagar, Kolkata - 700064, India}
\begin{abstract}
The multiplicity, finite system size and collision energy dependence of heat capacity ($C_V$), conformal symmetry breaking  measure (CSBM) and speed of sound ($c_s$) have been investigated using ALICE data for $p+p$ collisions at  $\sqrt{s}$ = 7 TeV.  The aim of this study is to ascertain the possibility of formation of a thermalized medium in such collisions. We find that there is a threshold in charged particle multiplicity beyond which $C_V$, CSBM and $c_s$ attain  plateau. The presence of such threshold in multiplicity is further reflected in the variation of these quantities with center-of-mass energy ($\sqrt{s}$). In order to have a grasp on experimentally obtained results, variation of average transverse momentum with multiplicity has also been studied. The experimental results have been contrasted with PYTHIA8 and it is 
found that PYTHIA8 is inadequate to explain the features reflected in these quantities, thereby indicating the possibility of thermalization in such small system. It is also observed that the finite size effects alone cannot explain the non-extensive nature of particle spectra in $p+p$ collisions. 
%\pacs{25.75.Dw,14.40.Pq}
\end{abstract}
\date{\today}
\maketitle
\section{Introduction}
\label{intro}
One of the main goals of relativistic heavy-ion collision experiments (RHICE) is to create and characterize quark gluon plasma (QGP). QGP is a deconfined state of quarks and gluons which can be realized at high density and temperature. It is expected that such high density and temperature can be created by colliding nuclei at relativistic energies. The characterization of QGP (i.e, determination of its equation of state, transport properties, etc.) can be done by analyzing experimental data  with the help of theoretical models. But the applicability of these theoretical models relies on certain assumptions.  For example, relativistic dissipative  hydrodynamics~\cite{Baier:2006um} can be used to analyze experimental data on anisotropic flow for the estimation of the viscous coefficients of QGP. Currently, it is not possible to prove from the first principle that the system produced in RHICE has achieved thermal equilibrium. Therefore, study on the validity  of the assumption of local thermalization (and hence applicability of hydrodynamics) with the help of experimental data is crucial. 
%The success of the hydrodynamical model depends on the understanding of the initial conditions 
%and equation of state required to solve the hydrodynamic equations.  

The assumption for the formation of partonic medium in RHICE is substantiated by experimental 
observations such as non-zero collective flow~\cite{Alver:2010rt}, suppression of $J/\psi$ yield~\cite{Kopeliovich:2011zz}, 
enhancement of strangeness~\cite{Agakishiev:2011ar}, suppression of high 
transverse momentum ($p_T$) hadrons~\cite{Isobe:2006vg}, etc. 
These observations have been  attributed to the formation of locally thermalized partonic medium which hydrodynamically evolves in space and time.
%The hydrodynamic modeling of RHICE can explain the anisotropic flow which has been linked 
%with the collective evolution, quantities
%like viscous coefficients of QGP has been extracted from such studies by using  
%the  experimental data from $p+p$ collisions as a benchmark with the assumption 
%that no partonic medium is formed in such collisions. 
Several parameters which are used to characterize the QGP formed in RHICE have been extracted
by analyzing experimental data where data from $p+p$ collision has been used as a benchmark. 
However, when small systems formed in high-multiplicity $p+p$ collisions show long-range ``ridge'' in two particle azimuthal 
correlations with a large pseudorapidity separation~\cite{Ridge1,Ridge2,Ridge3,Ridge4,Ridge5,Zhao:2017rgg}  imitating 
collectivity, then the role of $p+p$ collisions as a 
benchmark becomes obscured (although it is shown that non-medium effects can 
explain the features~\cite{MarkMac:2019plb,Braun:2015eoa}).  
In this regard, it is important to understand the role  of the number of constituents for the formation of the QCD medium and $p+p$ 
collisions can serve as a platform to address such issues. 
The multiplicity serves as a proxy to the number of constituents in a system formed in $p+p$ collisions. 
It may be recalled that the issue of thermalization in small systems was studied in 1953  by Landau~\cite{Landau:1965cpl} 
and in 1982, van Hove investigated thermalization and quark-hadron phase transition in proton-antiproton collisions using variation of average $p_T$ 
($\langle p_T\rangle$)
with multiplicity~\cite{VanHove:1982vk}. Recently, QCD thermodynamics in $p+p$ collisions has been
studied in~\cite{Hirsch:2018pqm}.

The purpose of this work is to investigate  - how some of the markers of thermalization {\it e.g.} the thermodynamic quantities like, $C_V$, CSBM and $c_s$ for small system vary with the quantities like
multiplicity, size and collision energy.  Because the chances for the system to achieve thermalization will
increase with the increase of these quantities. Therefore, any change in the variation of $C_V$, CSBM and $c_s$  with multiplicity (say) which is different from the change obtained from Monte-Carlo generators, which is devoid of thermalization (PYTHIA8 here) will 
signal on the possibility of thermalization. These quantities have been chosen because $C_V$ is one 
of the most basic and commonly used quantities which records the response of the system subjected to temperature stimulus. 
Similarly, $c_s$ provides the information on the equation of state of a thermal medium and the CSBM,
which can be expressed in terms of energy density ($\epsilon$), pressure ($P$) and temperature ($T$)
as $\text{CSBM}=(\epsilon-3P)/T^4$ (see~\cite{tracanomaly1,traceanomaly2} for details).  
In this context the variation of $\langle p_T\rangle$ of the hadrons  
with multiplicity connected to the temperature and
entropy of a thermal system respectively will also be examined. 
As there is no way to directly probe,
the spectra of produced hadrons are used to gain insight about the possible partonic phase. The ALICE data for $p+p$ collisions at $\sqrt{s}=7$ TeV have been used to obtain $C_V$, CSBM and $c_s$ and the results have been contrasted with PYTHIA8. 
The analysis using PYTHIA8 shows some degree of success in explaining some of the observations made in $p+p$ and p+Pb collisions, such as saturation of $\langle p_T \rangle$ of $J/\psi$ ~\cite{Thakur:2019qau, jpsiSat:2018} and that of charged particles~\cite{chargmpt}, as a function of charged particle multiplicity~\cite{Nature}.
Though variation of heat capacity with collision energy has been investigated~\cite{Basu:2016ibk,Li:2007ai} through temperature fluctuations for systems formed in RHICE, we are not aware of any studies in literature similar to the present one for small systems formed in $p+p$ collisions for understanding thermalization.

The paper is organized as follows. Formalism used in this work has been presented in section~\ref{formalism}.
Section \ref{Methodology} is devoted to discuss the methodology used for event generation 
in PYTHIA8. Results and discussions are presented in section ~\ref{result}. 
Finally in section \ref{sum}, we present the summary of our results.

\section{Formalism}
\label{formalism}
We will recall some of
the well-known thermodynamic relations in this section.
The system formed at the LHC energies at the central rapidity region 
will be dominated by gluons, which neither carry electric nor baryonic 
charges. Such a system can be described by one single thermodynamic variable, 
the temperature ($T$). 

Now we would like to quote the  standard thermodynamic expressions~\cite{Reif} for  
$C_{V}$, $c_s^2$ and 
entropy density (\textit s) below for a system with vanishing chemical potential
as:

\bea
C_{V}= \biggl(\frac{\partial \epsilon}{\partial T}\biggr)_{V},
\label{eq1}
\eea

\bea
s = \biggl(\frac{\partial P}{\partial T}\biggr)_{V},
\label{eq2}
\eea

\bea
c_s^2= \biggl(\frac{\partial P}{\partial\epsilon}\biggr)_{s}=s/C_{V},
\label{eq3}
\eea
where $V$ is the volume of the system. Now it is clear that to estimate the thermodynamic
quantities of our interest we need to know $\epsilon$, $P$, $s$, etc and
these quantities can be calculated  by using the phase space distribution functions
($f(E)$)~\cite{KolbTurner}. 
Interestingly, $f(E)$  for different hadrons can be measured experimentally by detecting their momentum 
distribution functions which allows us to connect data  with $C_V$, $c_s$, CSBM etc.  
%%%%%%%%%%%%%%%%%%%%%%%%%%%%%

In the present work the Tsallis non-extensive statistics~\cite{Cleymans:2014woa} is 
used to reproduce
the $p_T$-spectra of hadrons at kinetic freeze-out~\cite{Tsallis:1987eu,Tsallis:2008mc,Tsallis:2009zex}. 
%Although a derivation of non-extensivity from kinetic theory is not yet available,
The Tsallis-Boltzmann (TB) distribution function~\cite{wilk1,wilk2,wilk3} 
has been widely used to describe the results  from RHICE.
The TB distribution is given by:

\bea
f(E) \equiv  \frac{1}{\exp_{q}(\frac{E}{T})}
 \label{eq4}
\eea

where, 
\begin{equation}
\label{expq}
\exp_{q}(x) \equiv
  \begin{cases}
    [1+(q-1)x]^{\frac{1}{q-1}}       & \quad \text{if }  x > 0\\
    [1+(1-q)x]^{\frac{1}{1-q}}      & \quad \text{if }  x \le 0\\
  \end{cases}
\end{equation}
where $x = E/T$, $E$ is the energy ($E=\sqrt{p^2+m^2}$), 
$p$ and $m$ are momentum and mass of the particle, respectively. 
It is important to note that in the  limit, $q \rightarrow 1$, Eq. \eqref{expq} reduces to the standard 
exponential function,
\begin{eqnarray*}
\lim_{q \to 1} \exp_q(x) \rightarrow \exp(x).
\end{eqnarray*}
$T$ and $q$ appearing in TB distribution are extracted by fitting experimental data on 
hadronic $p_T$-spectra with this distribution. The parameter $q$ is called the non-extensive parameter which is a measure of degree of deviation from Boltzmann-Gibbs (BG) statistics. 
%The  $q$ can be determined from the dynamics of the system in principle~\cite{Tsallis:2009zex,CT2}. 
%Mathematically $q$ can vary from $0$ to $\infty$, although from the phenomenological point of view  it has been found that $q>1$ and thermodynamical considerations show that its upper bound is $4/3$~\cite{ts0}.
%It is possible for a few physical systems to determine $q$ analytically (see ~\cite{ts1}, \cite{ts2}, \cite{ts3}, \cite{ts4}). For systems formed in RHICE $q$ is treated as  free parameter and estimated by fitting transverse momentum spectra of hadrons. The sensitivity of the value of $q$ on multiplicity for different collision systems  ($p+p$, $p+A$ and $A+A$, $A$ stands for nucleus) have been investigated in~\cite{gbiro}
and $T$ appearing in this formalism obeys the fundamental 
thermodynamic relation:
\begin{equation}
T = \left.\frac{\partial U}{\partial S}\right|_{N,V} ,
\end{equation}
where $U$ is the internal thermal energy, $S$ is the total entropy $(=sV)$, 
$N$ is the number of particles and hence, the parameter $T$ can be called temperature, 
even though the system obeys the Tsallis and not the BG statistics.

The calculation of the number density ($n$),  energy density ($\epsilon$), pressure ($P$)
from the thermal phase space density, $f(E)$  is straight forward~\cite{KolbTurner}. These are given by,
$n=g/(2\pi)^3\int d^3pf(E)$, $\epsilon= g/(2\pi)^3\int d^3pEf(E)$ and $P= g/(2\pi)^3\int d^3p\frac{p^2}{3E}f(E)$. 
Analogously $\epsilon$, $P$, etc can be estimated  
for TB distribution by inserting $f(E)$ from Eq.~\ref{eq4}. 
The expression for energy density ($\epsilon$) then reads as \cite{Cleymans:2012ya,Kakati:2017xvr}:
\bea
\begin{aligned}
\epsilon &= \frac{g}{2\pi^2 }\int dp ~p^2 \sqrt{(p^2 + m^2)}\\
                    & \times [1+\frac{(q-1)\sqrt{(p^2 + m^2)}}{T}]^\frac{-q}{q-1},
\label{eq5}
\end{aligned}
\eea
where, $g$ is the degeneracy factor.

Similarly the expression for pressure ($P$) is given by~\cite{Cleymans:2012ya,Kakati:2017xvr}:
\bea
\begin{aligned}
P &= \frac{g}{2\pi^2 }\int dp ~p^4\frac{1}{3\sqrt{(p^2 + m^2)}}\\
        & \times [1+\frac{(q-1)\sqrt{(p^2 + m^2)}}{T}]^\frac{-q}{q-1},
\label{eq6}
\end{aligned}
\eea

The expression for $C_V$ as given in Eq.~\ref{eq1}  can be obtained  from Eq.~\ref{eq5}  as: 

 \bea
  \begin{aligned}
  C_{V} &= \frac{qg}{2\pi^2 T^2}\int dp ~p^2 (p^2 + m^2)\\
               & \times [1+\frac{(q-1)\sqrt{(p^2 + m^2)}}{T}]^\frac{1-2q}{q-1},
 \label{eq7}
  \end{aligned}
\eea
The dimensionless quantity $I/T^4$, where $I=\epsilon-3P$  called
the trace anomaly~\cite{tracanomaly1,traceanomaly2}
or CSBM can be expressed as:
\bea
\begin{aligned}
\frac{I}{T^4} &= \frac{g}{2\pi^2 T^4}\int dp~ p^2 \sqrt{(p^2 + m^2)}\\  
              & \quad{}[1-\frac{p^{2}}{(p^2+m^2)}] \\
              & \times {[1+\frac{(q-1)\sqrt{(p^2 + m^2)}}{T}]^\frac{-q}{q-1}},                                            
\end{aligned}
\label{eq8}
\eea
The squared velocity of sound ($c_s^{2}$) in QGP is given by:
\bea
c^2_s =\frac{\frac{gq}{6\pi^2 T^2}\int dp~ p^4 \times [1+\frac{(q-1)\sqrt{(p^2 + m^2)}}{T}]^\frac{1-2q}{q-1}}{C_{V}}
  \label{eq9}
\eea
and finally, the $\langle p_{T} \rangle$ for 
the TB  distribution can be estimated from the following expression:
\bea
\langle p_T \rangle =\frac{\int dp_{T}~p_{T}^{2}{f(E)^{q}}}{\int dp_{T}~p_{T}{f(E)^{q}}}.
  \label{eq_extra}
\eea
{\it i.e.}
\bea
\langle p_T \rangle =\frac{\int dp_{T}~p_{T}^{2}[1+(q-1)\frac{\sqrt{(p_{T}^{2}+m^{2})}}{T}]^{\frac{-q}{q-1}}}{\int dp_{T}~p_{T}[1+(q-1)\frac{\sqrt{(p_{T}^{2}+m^{2})}}{T}]^{\frac{-q}{q-1}}}.
  \label{eq10}
\eea
\section{Event generation and Analysis Methodology}
\label{Methodology} 
In order to make a comparative study, results obtained in this work are compared with QCD-inspired Monte-Carlo generator PYTHIA8, which is a amalgam of various physics mechanisms like hard and soft interactions, initial and final-state parton showers, fragmentation, multipartonic interactions, color reconnection, rope hadronization etc~\cite{Sjostrand:2006za}. This model is used 
here to simulate $p+p$ collisions at ultra-relativistic energies. Detailed explanation on PYTHIA8 physics processes and their implementation can be found in Ref.\cite{PYTHIA8html}.

We have used 8.215 version of PYTHIA, which includes multi-partonic interaction (MPI). MPI is crucial to explain the underlying events multiplicity distributions. Also, this version includes color reconnection which mimics the flow-like effects in pp collisions~\cite{Ortiz:2013yxa}. 
It is crucial to mention here that PYTHIA8 does not have in-built thermalization. However, as reported in Ref. \cite{Ortiz:2013yxa}, the color reconnection (CR) mechanism along with the multi-partonic interactions (MPI) in PYTHIA8 produces the properties which mimics  thermalization of a system such as radial flow and mass dependent rise of mean transverse momentum. 
 Apparently the PYTHIA8 with MPI and CR has the ability to produce the features 
similar to thermalization.

QCD processes in PYTHIA8 are categorised as soft and hard QCD processes, where production of heavy quarks are included in the latter. We have simulated the inelastic, non-diffractive component of the total cross-section for all the soft QCD process (SoftQCD:all = on) and Hard QCD process (HardQCD:all = 0) separately. MPI based scheme of color reconnection (ColorReconnection:reconnect =0) are also included. 
We have generated 100 million events with 4C tune (Tune:pp=5)~\cite{Corke:2010yf}, which give sufficient statistics to obtain $p_{T}$-spectra even in high-multiplicity events. To check the compatibility of tunes used in this work, we have compared simulated results obtained from hard and soft QCD tune of PYTHIA8 with the experimental data~\cite{Aad:2010ac} as shown in Fig.~\ref{data_model_ATLAS}. Here, we have compared PYTHIA8 simulated data with ATLAS data~\cite{Aad:2010ac} as at the time of this work, there is no mini-biased ALICE data available for transverse momentum distribution of charged particles in $p+p$ collisions at $\sqrt{s}$=7 TeV. The motivation to contrast the PYTHIA8 generated results with the experimental data is to  show that the soft processes fit the data reasonably well as shown in ~Fig~\ref{data_model_ATLAS}. This comparison makes it clear that softQCD tune of the PYTHIA8 is suitable for the present work.

The generated events are categorised into seven multiplicity bins as (0-2), (2-4), (4-8), (8-11), (11-14), (14-18), (18-24) from which charged-particle pseudorapidity densities ($\langle dN_{\rm ch}/d\eta\rangle$) at mid-rapidity are obtained. The $p_T$ distribution generated by PYTHIA8 for different multiplicity bins are now fitted with the following expression with $T$ and $q$ as fitting parameter~\cite{Li:2015jpa}: 

%The generated events are categorized into ten multiplicity bins.  The charged-particle pseudorapidity densities ($dN_{\rm ch}/d\eta$) in an event at mid-rapidity in different multiplicity classes are (0-2), (2-4), (4-8), (8-11), (11-14), (14-18), (18-24). The $p_T$ distribution generated by PYTHIA*for different multiplicity bins are now fitted with the following expression with $T$ and $q$ as fitting parameter~\cite{Li:2015jpa}: 
\begin{eqnarray}
\label{eq12}
\left.\frac{1}{p_T}\frac{d^2N}{dp_Tdy}\right|_{y=0} = \frac{gVm_T}{(2\pi)^2}
\left[1+{(q-1)}{\frac{m_T}{T}}\right]^{-\frac{q}{q-1}},
\end{eqnarray}
where $m_{\rm T} = \sqrt{p_T^2 + m^2}$. 
The fitting parameters, $T$ and $q$ depends on the mass of the hadrons.

The $p_T$-spectra of $\pi^{\pm}$, $K^{\pm}$, $K^{*0} + \overline{K^{*0}}$ and $p + \overline{p}$ 
from simulated data at the mid-rapidity ($|\eta| < 0.5 $) for different 
multiplicity bins in $p+p$ collisions at $\sqrt{s}$ = 7 TeV
have been considered. The fitting of the PYTHIA8 generated spectra by Eq.~\ref{eq12} is 
displayed in Fig.~\ref{fit_Tsallis_pythia}. Figure ~\ref{chi_square_pythia} shows the quality of 
fitting in terms of $\chi^{2}/NDF$ as function of multiplicity  which shows that the quality of fitting is reasonably good for all the particles under consideration at all 
multiplicity classes except for $p + \overline{p}$ at low multiplicity class.

Figures~\ref{T_vs_nch_pythia_ALICE} and ~\ref{q_vs_nch_pythia_ALICE} show the comparison 
of the parameters $T$ and $q$ extracted from the experimental data and the PYTHIA8 generated 
results for different charged particle multiplicities~\cite{Thakur:2016boy,Khuntia:2018znt}. 
%Figure~\ref{T_vs_nch_pythia_ALICE} shows that the temperature parameter for $\pi^{\pm}$ obtained from experimental data are comparable to the PYTHIA8 simulated values in the low multiplicity region but it becomes smaller in the higher multiplicity region. For $K^{*0} + \overline{K^{*0}}$, experimental data seem to have higher $T$  than simulated data. However, $K^{\pm}$ and $p+\overline{p}$, simulated results clearly overestimates the experimental data. Non-extensive parameter ($q$) as shown in Fig.~\ref{q_vs_nch_pythia_ALICE} seems to have comparable value within uncertainties both for simulated results and experimental data.

%Further, the space-time evolution of hadronic and heavy-ion collisions at the LHC energies could be thought of following a cosmological expansion of the produced fireball. In this scenario, as the fireball expands and cools down, it leaves with a temperature profile with time. Different identified particles decouple from the fireball giving the signature of a mass dependent particle freeze-out -- higher mass particles decoupling from the system earlier in time. In this work, we have considered such a scenario and have evaluated various thermodynamic quantities at different decoupling points of final state particles from the
%produced fireball.

%With the detailed analysis methodology and (T, q) values obtained from PYTHIA8, we now move to discuss the results in the next section.
With the detailed analysis methodology and (T, q) values obtained from PYTHIA8, we now move to discuss the results obtained by comparing ALICE experimental data and simulated data in the next section.

\bef[ht]
\bc
\includegraphics[scale=0.38]{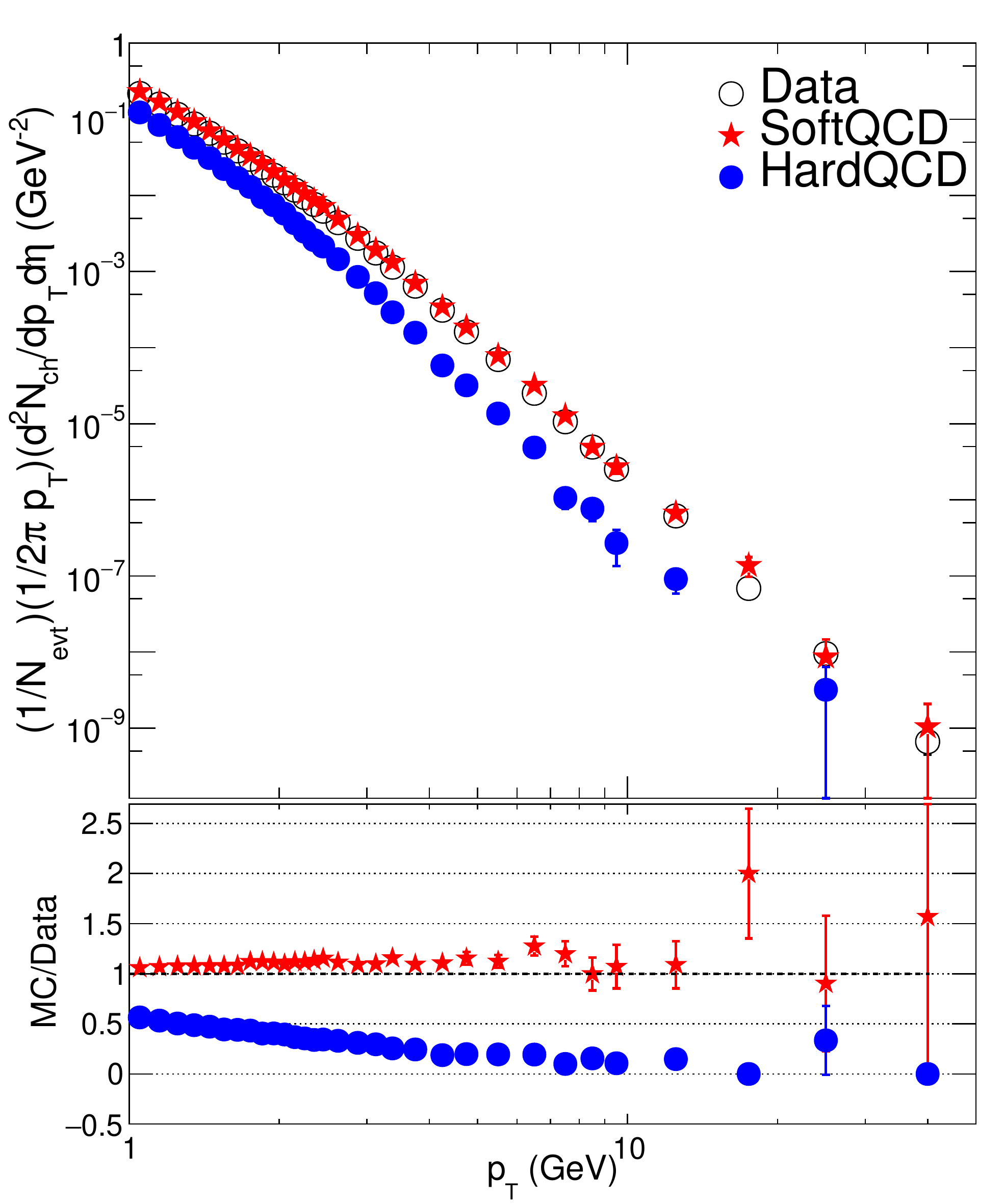}
\caption  {(Color online) The upper panel shows the comparison of experimental data~\cite{Aad:2010ac}, HardQCD and SoftQCD tunes of PYTHIA8 for p+p collisions at $\sqrt{s}$ = 7 TeV. The black open circles are experimental data, red stars and blue solid circles are PYTHIA8 
simulated data with SoftQCD and HardQCD tunes, respectively. 
The lower panel shows the ratio of PYTHIA8 to experimental data for both softQCD and hardQCD cases.
The vertical lines indicate the error bars.}
 \label{data_model_ATLAS}  
 \ec
 \eef

\bef[ht]
\bc
\includegraphics[scale=0.445]{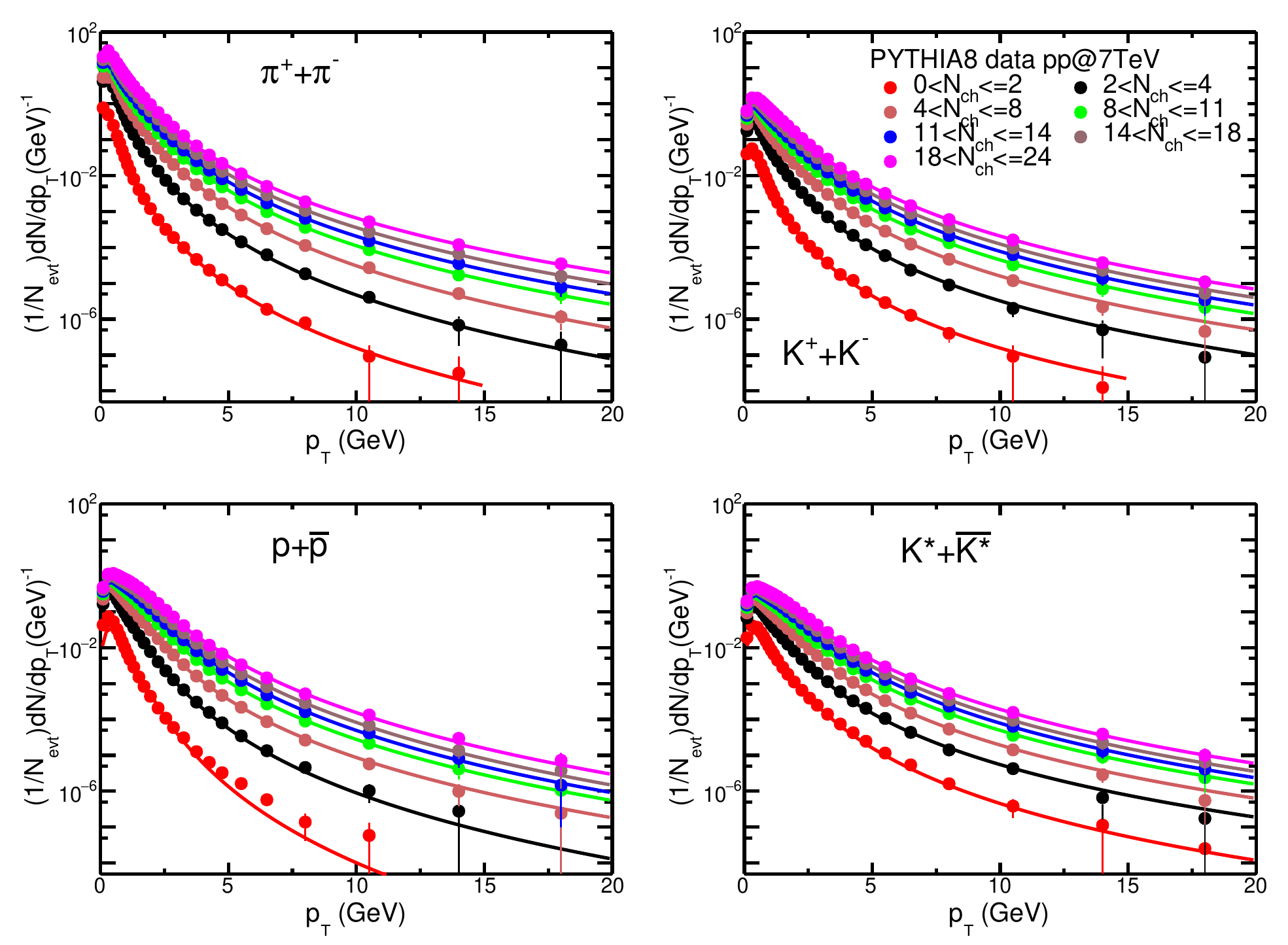}
\caption  {(Color online) Fitting of PYTHIA8 generated $p_{\rm T}$-spectra of $\pi^{\pm}$, $K^{\pm}$, $K^{*0} + \overline{K^{*0}}$ and $p + \overline{p}$ using Tsallis distribution (Eq.~\ref{eq12}) 
for various multiplicity classes at mid-rapidity for $p+p$ collisions at $\sqrt{s}$ = 7 TeV.
In the legend of the figure, we have used a short notation $N_{\rm ch}$ 
for $\langle dN_{\rm ch}/d\eta\rangle$.}
 \label{fit_Tsallis_pythia}  
 \ec
 \eef

 \bef[ht]
\bc
 \includegraphics[scale=0.40]{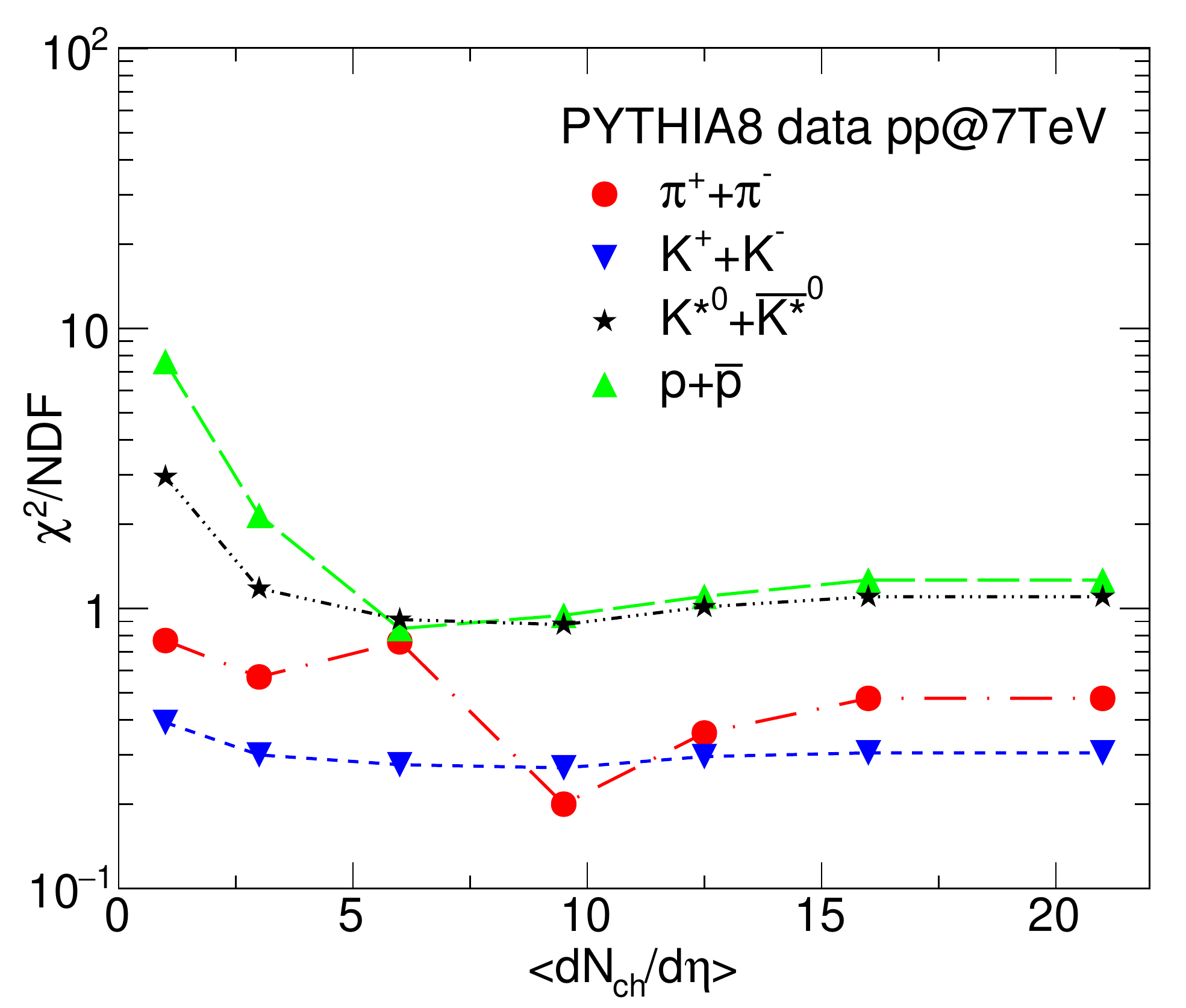}
\caption  {(Color online) $\chi^{2}$/NDF for $\pi^{\pm}$, $K^{\pm}$, $K^{*0} + \overline{K^{*0}}$ and $p + \overline{p}$ as a function of charged particle multiplicity obtained 
by fitting Tsallis distribution
(Eq.~\ref{eq4}).}
 \label{chi_square_pythia}  
 \ec
 \eef

 \bef[ht]
\bc
 \includegraphics[scale=0.446]{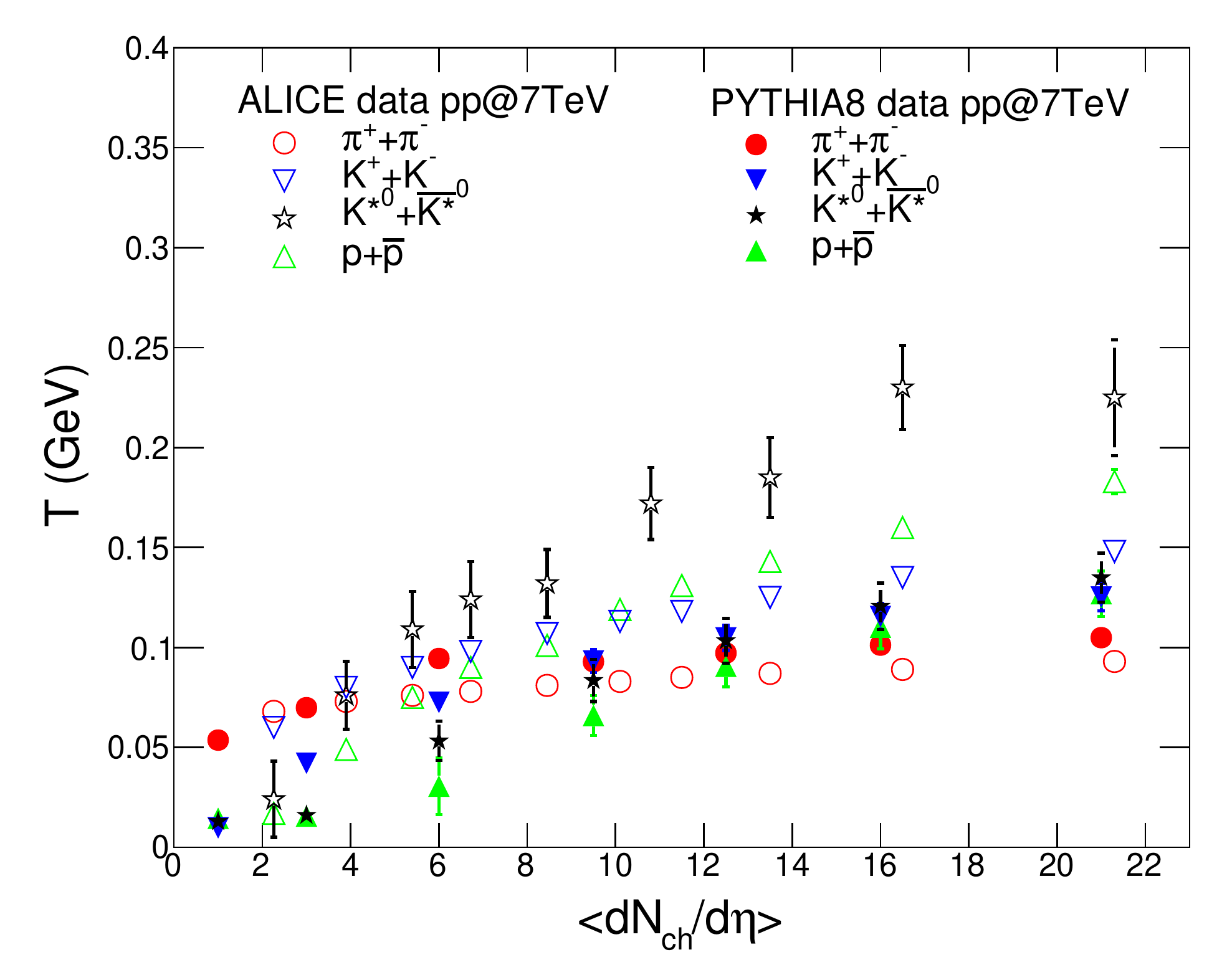}
 \caption  {(Color online) Multiplicity dependence of T for $p+p$ collisions at $\sqrt{s}$ = 7 TeV 
obtained by using Eq.~\ref{eq12} as a fitting function for the PYTHIA8 simulated numbers (solid markers) and 
experimental data (open markers)~\cite{Khuntia:2018znt}.}
\label{T_vs_nch_pythia_ALICE}  
 \ec
 \eef

 \bef[ht]
\bc
 \includegraphics[scale=0.446]{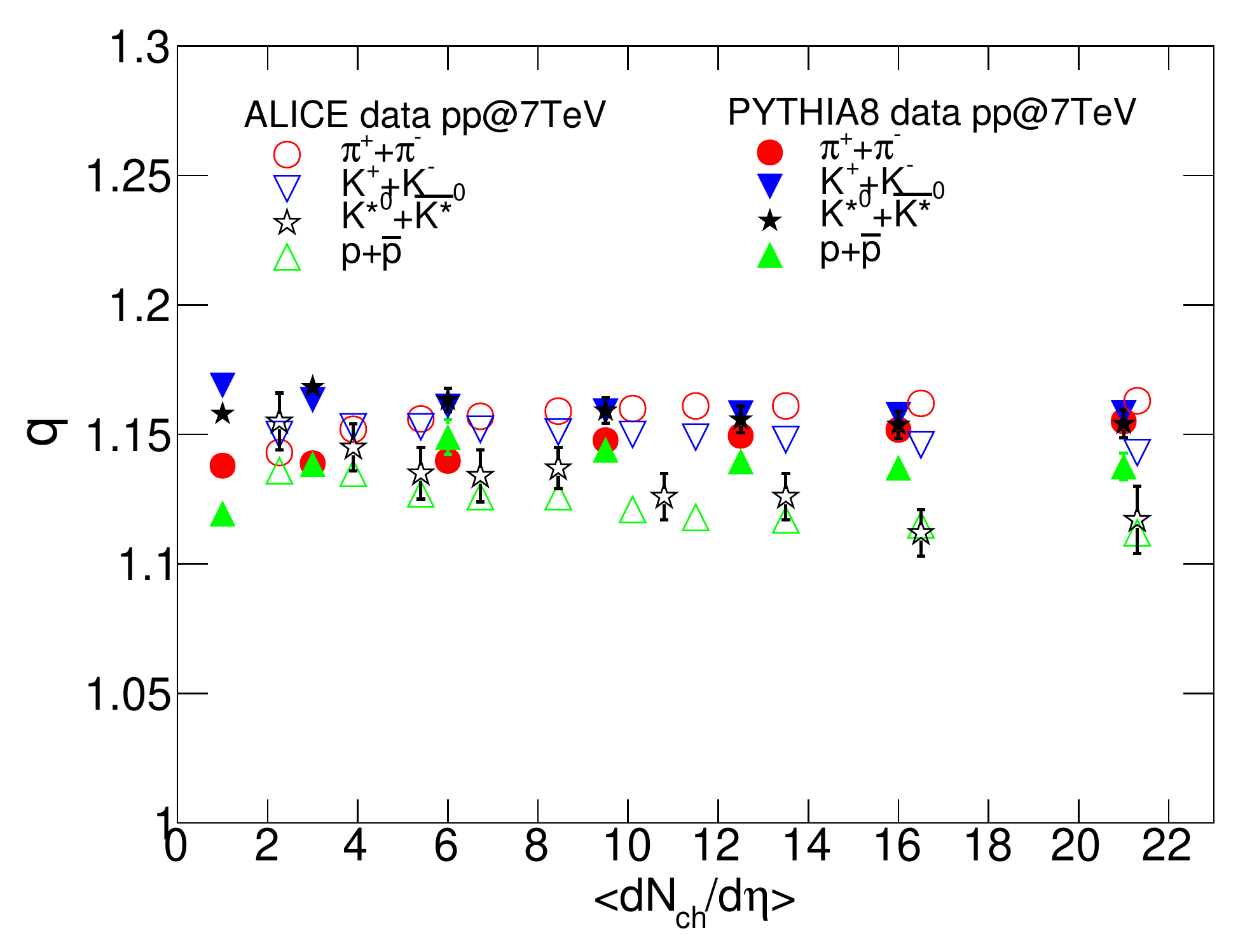}
 \caption  {(Color online) 
Same as Fig.\ref{T_vs_nch_pythia_ALICE}  for variation
of the parameter $q$ with charged particle multiplicity.}
 \label{q_vs_nch_pythia_ALICE}  
 \ec
 \eef

 \section{Results and Discussions}
 \label{result} 
The thermodynamic quantities, $C_V$, CSBM, $c_s$ and $\langle p_T\rangle$ can be estimated by using 
Eqs.~\ref{eq7}, ~\ref{eq8}, ~\ref{eq9}, and ~\ref{eq10} with the values of $T$ and $q$ extracted by parameterizing the $p_T$ spectra of identified hadrons using TB distribution. We note that a similar kind of approach has been used to study the variation of $C_{V}$ with $\sqrt{s_{NN}}$ at the freeze-out surfaces in the context of heavy-ion collisions~\cite{Basu:2016ibk}. The  ALICE data on the $p_T$-spectra originating from $p+p$ collisions at 7 TeV collision energy~\cite{Acharya:2018orn} have been used to extract the values of T and q for each multiplicity in Ref.~\cite{Khuntia:2018znt}. 
%In the present work, we have used $T$ and $q$ from Ref.~\cite{Khuntia:2018znt} for calculating the thermodynamic quantities expressed in previous section. 
It is found that the values of $T$ and $q$ depend on hadronic species hinting at different decoupling or freeze out temperature for different hadrons~\cite{Thakur:2016boy}. In general, the hadrons with higher inverse slope (of $p_T$-spectra) is expected to come either from early stage and/or 
suffer more transverse flow. In the present study, we consider hadronic spectra of pion ($\pi^{\pm}$), 
kaon ($K^{\pm}$), neutral kstar ($K^{*0} + \overline{K^{*0}}$) and proton ($p + \overline{p}$).  

The variation of $C_{V}$, $C_{V}/\langle n_{i} \rangle$, (where i=$\pi^{\pm}$, $K^{\pm}$, 
$K^{*0}+\overline{K^{*0}}$ and $p+\overline{p}$), $C_{V}/T^3$, $C_{V}/(\epsilon+P)$, 
CSBM, $c_s^2$ and $\langle p_{T} \rangle$ with  
$\langle dN_{\rm ch}/d\eta\rangle$ at mid-rapidity have been considered. 
Here $<n_{i}>$ (in $\rm GeV^{3}$) is the number density (number per unit volume)   
of the hadron $i$,  obtained by integrating Eq.~\ref{eq4} over three momentum and  
$(\epsilon+P)$ is the enthalpy density. The values of  $\langle dN_{\rm ch}/d\eta \rangle$ obtained in 
the experiment for different multiplicity classes tabulated in table
~\ref{table:mult_info}  
(see Ref.~\cite{Acharya:2018orn} for details). 
We also investigate whether finite system size alone can account for 
non-extensivity observed in the spectra. To make a distinction between systems with and without  
thermalization we contrast the results with the PYTHIA8 simulated outputs under the same collision 
condition. 

\begin{table*}[htbp]
\caption[p] {Average charged-particle pseudorapidity densities corresponding to different event multiplicity classes \cite{Acharya:2018orn}.}
\label{table:mult_info}
\begin{tabular}{c|c|c|c|c|c|c|c|c|c|c|c|}
\hline
\multicolumn{2}{|c|}{${\bf Class ~name}$}&Mul1&Mul2&Mul3&Mul4&Mul5&Mul6&Mul7&Mul8&Mul9&Mul10\\
\hline
\multicolumn{2}{|c|}{ $  \bf \big<{\frac{dN_{ch}}{d\eta} } \big>$} &21.3$\pm$0.6&16.5$\pm$0.5&13.5$\pm$0.4&11.5$\pm$0.3&10.1$\pm$0.3&8.45$\pm$0.25&6.72$\pm$0.21&5.40$\pm$0.17&3.90$\pm$0.14&2.26$\pm$0.12\\
\hline
\end{tabular}
\end{table*}

\subsection{Multiplicity dependence of heat capacity}
\label{mult_C_{V}}
As mentioned before, $C_V$ is one of the most fundamental quantities that gives the response of a 
thermal system under the influence of temperature stimulus.  
It gives the measure of how variation of temperature changes the entropy 
of a system ($\Delta S = \int \frac{C_{V}}{T} dT$). The change in entropy is a 
good observable for studying the phase transition. In the context of heavy-ion collisions, entropy ($S$) per unit rapidity ($y$),
$dS/dy$ can be connected to the  the corresponding multiplicity ($dN/dy$).
Therefore, the heat capacity acts as bridging observable for experimental measurement and theoretical models. 
For a strongly interacting system sufficient heat energy should be supplied to 
overcome the `binding force' caused by the interaction to increase the temperature
{\it i.e.} to supply
adequate randomized kinetic energy by the constituents of the system.
In other words, the mechanism of randomization to increase the temperature will require supply of more heat energy for 
strongly interacting system compared to that needed for the weakly interacting system. That is heat energy 
supplied to the strongly interacting system will not be entirely utilized to increase the 
temperature, some amount will be used to weaken the binding. Hence the increase in temperature in a strongly interacting system 
will be less than a weakly interacting system for a given amount of energy supplied to the system. 
Thus, the heat capacity bears the effects of strength of interaction among constituents of the system and represents 
the ease of randomization for the particular phase of the matter. Therefore,
for weakly interacting gas increase of temperature has negligible
effects on change in interaction strength. As a result
its scaled value, $C_{V}/\langle n_{i} \rangle$ will display a plateau. 
%In general, if some conditions cause changes in the strength of interaction, 
%then ease of randomization, and hence, 
%the heat capacity will change with that condition. 
This makes heat capacity a good observable to study how correlation 
and randomization competes in the system. The variation of heat with multiplicity in $p+p$ collision 
gives opportunities to better understand the randomization and the change in the 
strength of correlation  with number of constituents in the QCD system. 

The variation of $C_{V}$ with $\langle dN_{\rm ch}/d\eta\rangle$ 
for $\pi^{\pm}$, $K^{\pm}$, $K^{*0} + \overline{K^{*0}}$ and $p + \overline{p}$ 
extracted from ALICE data has been displayed in Fig.~\ref{fig1}. The result has been contrasted with the output obtained from PYTHIA8 simulation 
at the same $p+p$ colliding energy. It is observed that results from PYTHIA8 which do not contain medium effects differ from data. 
Also, it is noted that the heat capacity increases with increase in multiplicity. 
If a thermalized medium is formed, then, in the ideal gas limit, heat capacity varies linearly  
with number of particles ($C_{V} \propto \langle n_{i} \rangle$). 
%From Fig.~\ref{fig1}, it is evident that the 
%lighter particles like $\pi^{\pm}$ and $K^{\pm}$ tend to show such proportionality trends, 
%however, the heavier particles like $K^{*0} + \overline{K^{*0}}$ and $p + \overline{p}$ show 
%deviation from such trend. This may be due to the fact that heavier hadrons either decouple from the 
%system earlier than the lighter hadrons in the course of evolution.

Therefore, in Fig.~\ref{fig2} we depict the variation of $C_{V}$ scaled by $\langle n_{i} \rangle$ 
extracted from ALICE data as well as PYTHIA8 as a function of $\langle dN_{\rm ch}/d\eta\rangle$ 
for $\pi^{\pm}$, $K^{\pm}$, $K^{*0} + \overline{K^{*0}}$ and $p + \overline{p}$  of  
ALICE data and results from PYTHIA8. It is observed that $C_{V}/\langle n_{i} \rangle$, (where i = $\pi^{\pm}$, $K^{\pm}$)
tend to almost saturate for high-multiplicity,  however, a slow variation is observed for 
$i=\overline{K^{*0}}$ and $p + \overline{p}$.
It is important to note that the pionic and kaonic matter (for $\langle dN_{\rm ch}/d\eta \rangle >$ 8) have approximately similar value of $C_V/\langle n_{i} \rangle$ for both experimental and MC data.

%%%%%%%%%%%%%%%%%%%%%%%%%%%%%%%%xyz

%Here, also, the explanation of the results displayed in Fig~\ref{fig1} can be brought 
%in as in the Boltzmann's limit for thermalized medium, $C_{V}/<n_{i}>$, should be constant 
%for constant temperature of the system irrespective of other macroscopic conditions like 
%number of particles that affects the many-body interaction environment. 
The observed saturation in specific heat in its variation with multiplicity can be attributed to the possibility of thermalization in the system.
%due to change in interaction environment with increasing multiplicity such that interaction 
%strength reduces with increase in number of particles. This may account for the observed 
%saturation, suggesting that with the increase in multiplicity the produced system goes towards 
%thermalization. 
We also notice that results from PYTHIA8 are not in good agreement with ALICE data for heavier particles like $K^{*0} + \overline{K^{*0}}$ and $p + \overline{p}$  . Here $\langle n_{i} \rangle$ has a fractional value in unit $\rm GeV^{3}$, this makes the value of $C_{V}/\langle n_{i} \rangle$ greater than $C_{V}$, as evident from the results displayed in 
Figs.~\ref{fig1} and Fig~\ref{fig2}.

%{\color{red}For massless ideal gas, number density (n) varies as $T^{3}$, 
%so $C_{V}/T^{3}$ should be a constant for a system of weakly 
%interacting massless particles}. 

Fig.~\ref{fig3} shows the variation of $C_{V}$ (scaled by $T^{3}$) with 
$\langle dN_{\rm ch}/d\eta \rangle$ for $\pi^{\pm}$, $K^{\pm}$, 
$K^{*0} + \overline{K^{*0}}$ and $p + \overline{p}$  extracted from  ALICE data and PYTHIA8.
It is observed that $C_{V}/T^3$ for $\pi^{\pm}$, $K^{\pm}$, 
$K^{*0} + \overline{K^{*0}}$  increases with multiplicity and display a saturation (within the error bars) when 
$\langle dN_{\rm ch}/d\eta\rangle > 8$, whereas  $p + \overline{p}$ displays an increasing trend with 
$\langle dN_{\rm ch}/d\eta \rangle$ without any sign of saturation. This may be a hint to the fact that lower mass 
particles like $\pi^{\pm}$, $K^{\pm}$ behave as weakly interacting thermalized particles beyond 
certain multiplicity, whereas heavier mass particles may not witness a thermalized medium. 
Here, also PYTHIA8 results are not in good agreement with ALICE data. 

Fluid dynamical equation in non-relativistic limit (Euler equation in the limit of small 
flow velocity ($v$) for ideal fluid) can be written as: $(\epsilon+P)\partial{\vec{v}}/\partial t
=-{\vec{\nabla}} P$, where $(\epsilon+P)$ is the enthalpy density. Comparison of this equation
with the non-relativistic classical mechanical equation of a particle moving with 
velocity $v$ in a potential, $\phi$: $m d\vec{v}/dt=-\vec{\nabla}\phi$,
indicates that enthalpy density plays the role of mass (inertia) in fluid dynamics. 

Since enthalpy density, $(\epsilon+P)$, acts as inertia for change in velocity 
for a fluid cell in thermal equilibrium, we display the change in $C_{V}$ scaled by 
enthalpy density as a function of multiplicity 
in Fig~\ref{fig4}. The saturations of  $C_{V}/(\epsilon+P)$ and $C_{V}/\langle n_{i} \rangle$ 
in their variations with multiplicity show an interesting trend in which, at the saturation region, corresponding values for all the particle species tend to converge. This means that with the increase in the number of particles the system 
achieved randomization.
This is expected when particles in the system evolves collectively
with common interaction environment.

The effects of non-extensive parameter, $q$ on heat capacity has 
been shown in Fig~\ref{fig5} through the ratio,  $c_{v}/{c_{v}}_{q\rightarrow 1}$, where, 
$c_v=C_V/<m_in_i>$, here, $<m_{i}n_{i}>$ is the mass density of the hadron $i$. 
It may be mentioned that the $C_{V}$ is obtained here by fitting the TB distribution to the ALICE data, therefore, $C_{V}$ depends on $q$. From Fig~\ref{fig5}, the ratio seems to approach toward saturation for
$\langle dN_{\rm ch}/d\eta \rangle >5$, implying that new environment of interaction is 
set-off after $\langle dN_{\rm ch}/d\eta\rangle \approx (4-6)$, however, the ratio does not approach 
unity  except for $K^{\pm}$. This indicates 
that the system has not achieved the state to be described by BG statistics.

%\vspace{0.08em}.eps
\bef[ht]
\bc
\includegraphics[scale=0.440]{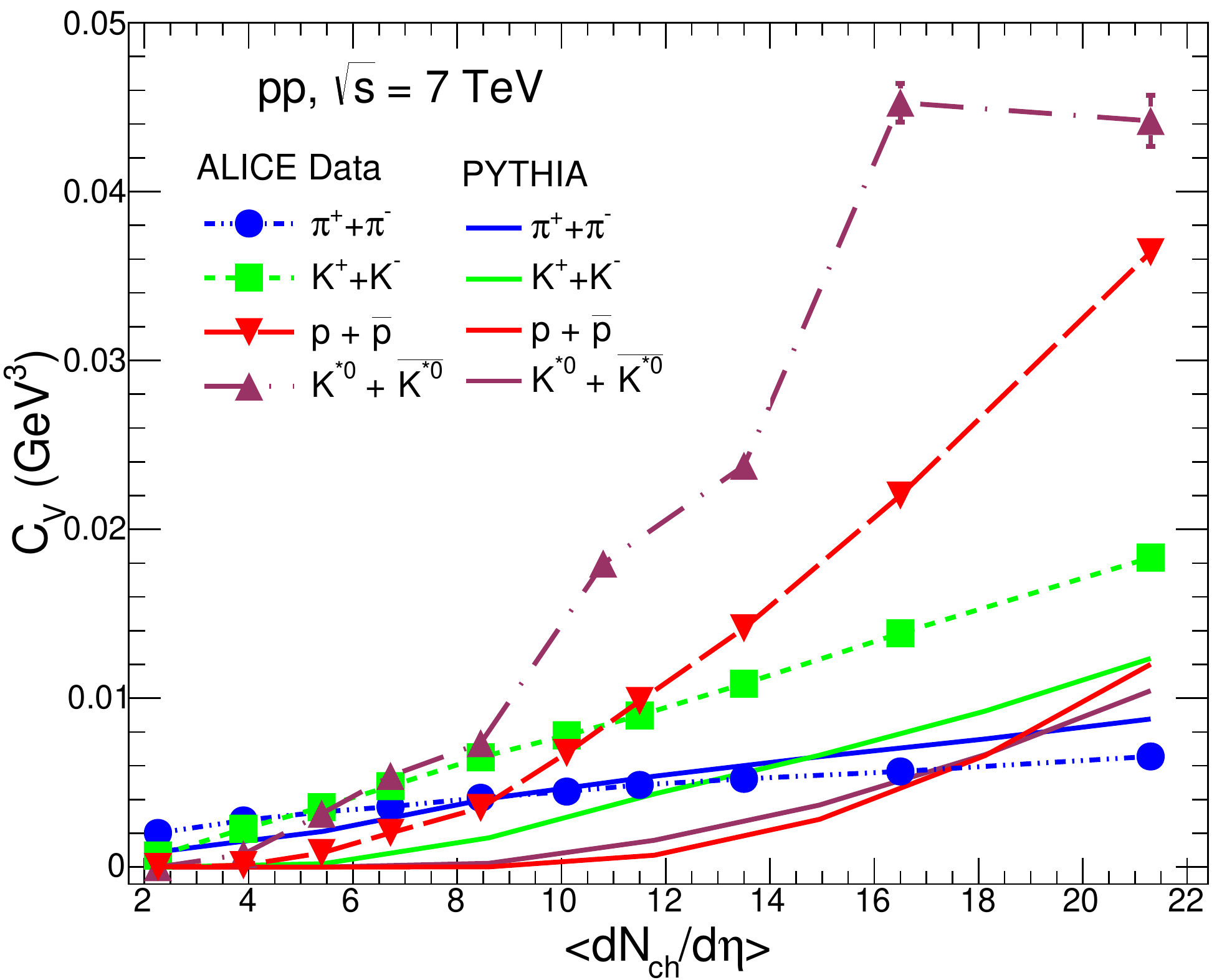}
\caption  {(Color online) Heat capacity obtained using TB distribution as a function of multiplicity. 
Dashed (solid) lines represent results obtained using ALICE (PYTHIA8 simulated) data, respectively for $p+p$ collisions at $\sqrt{s}$ = 7 TeV.}
\label{fig1}  
\ec
\eef

 \bef[ht]
 \bc
 \includegraphics[scale=0.440]{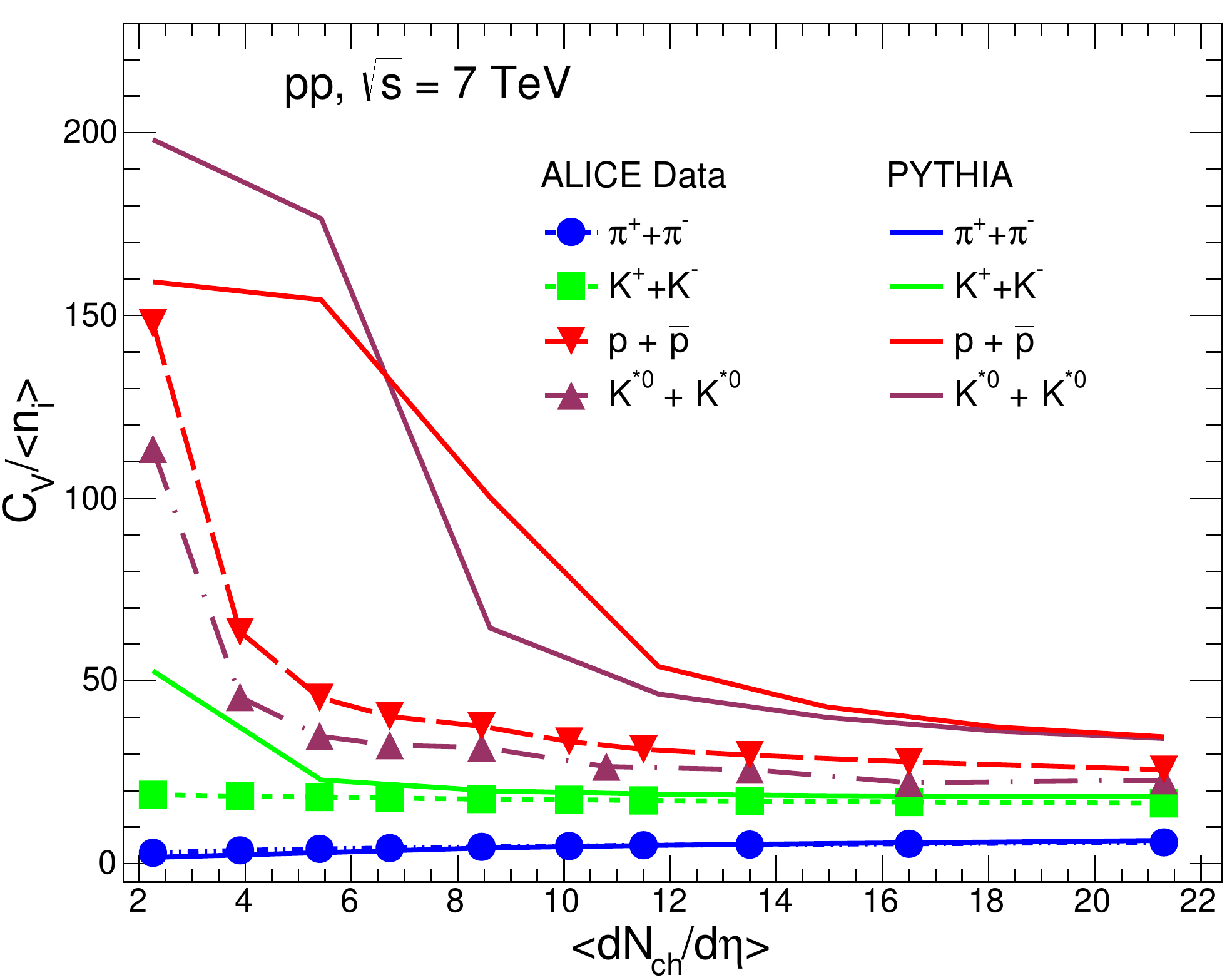}
 \caption  {(Color online) 
Same as Fig.~\ref{fig1} showing the variation of $C_V/<n_i>$ 
with charged multiplicity.}
\label{fig2} 
 \ec
 \eef

 \bef[ht]
 \bc
 \includegraphics[scale=0.440]{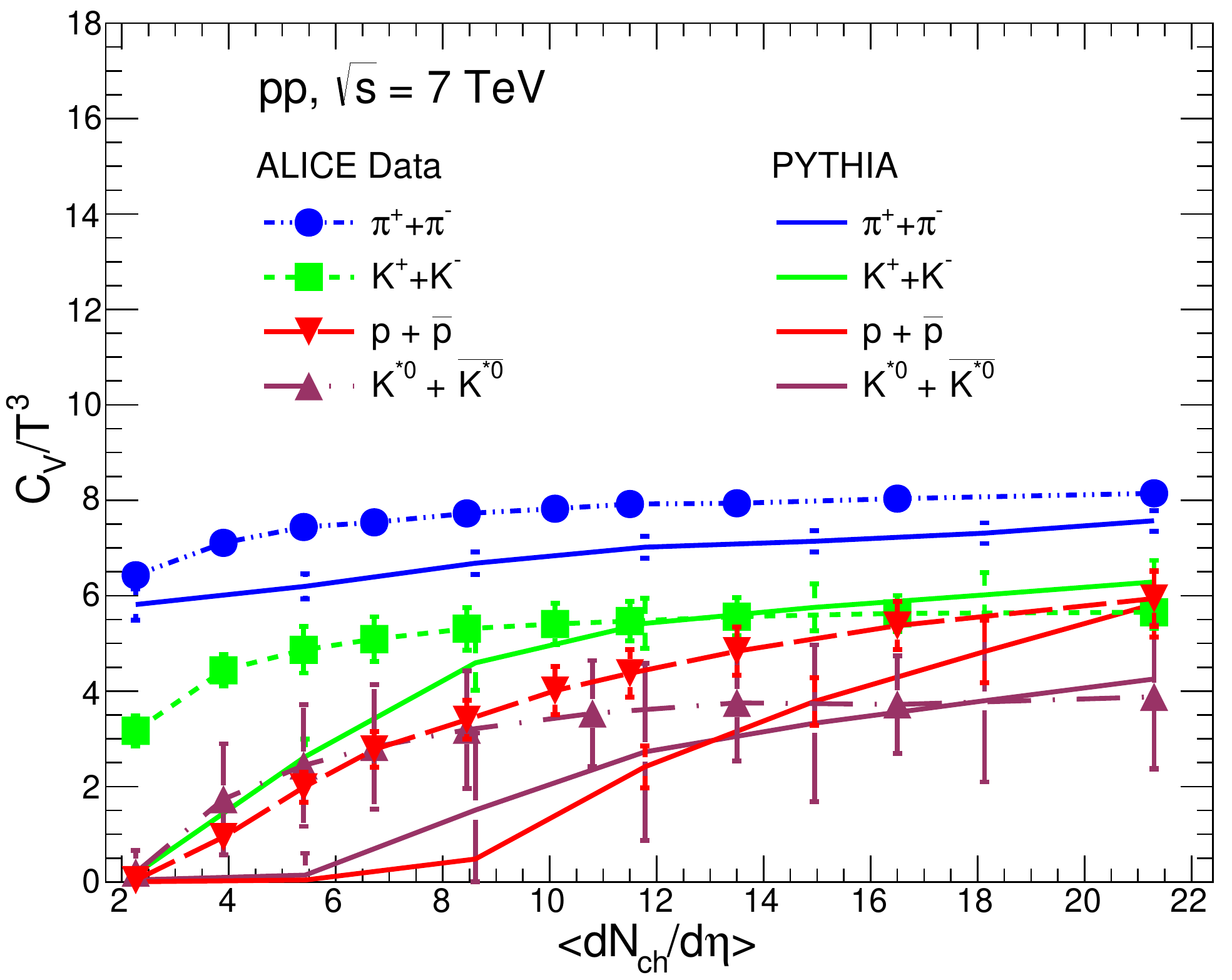}
 \caption  {(Color online) 
Same as Fig.~\ref{fig1} showing the variation of $C_V/T^3$ 
with charged multiplicity.}
\label{fig3} 
 \ec
 \eef

 \bef[ht]
 \bc
 \includegraphics[scale=0.445]{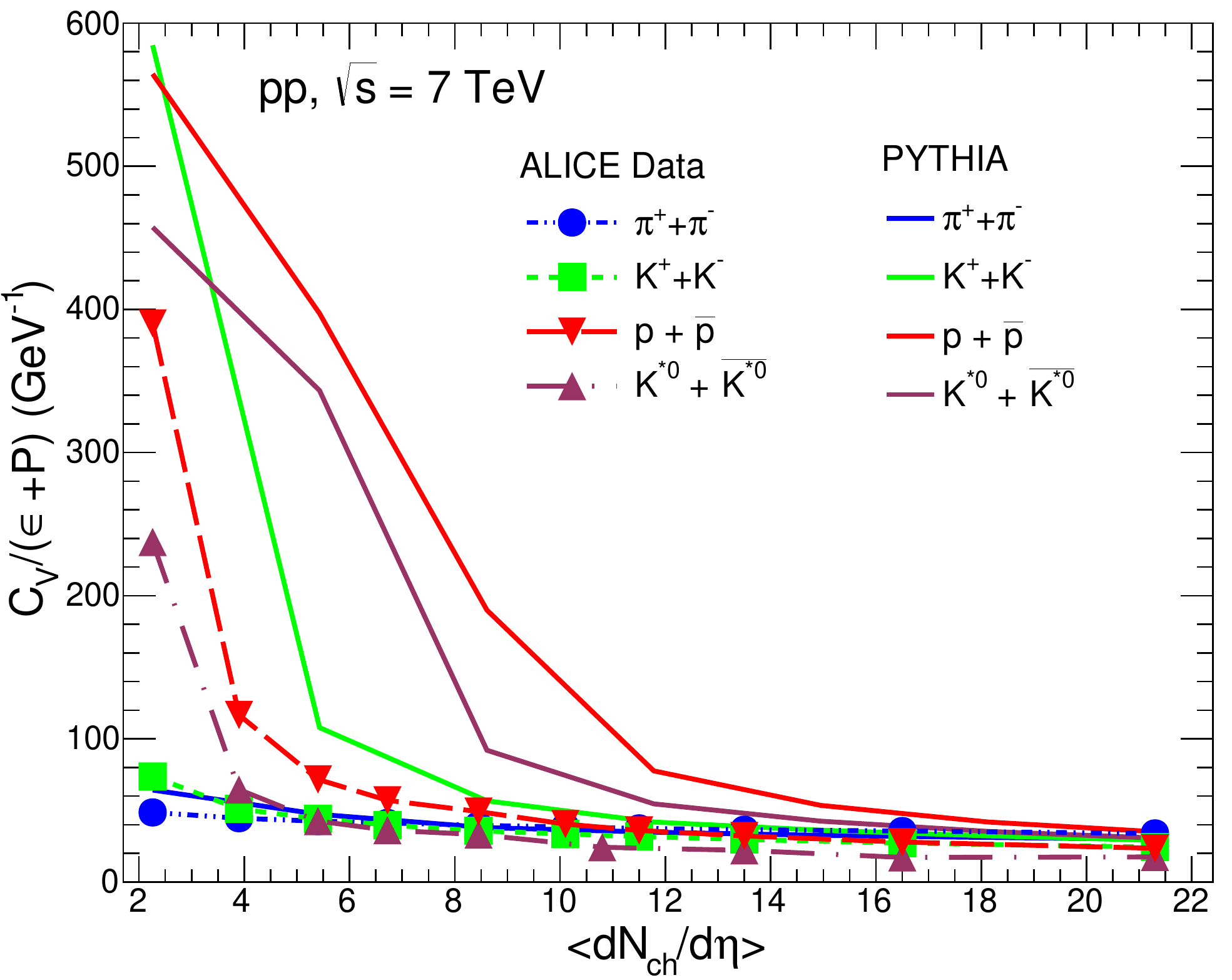}
 \caption  {(Color online) 
Same as Fig.~\ref{fig1} showing the variation of $C_V/(\epsilon+P)$ 
with charged multiplicity.}
\label{fig4} 
 \ec
 \eef 

\bef[ht]
 \bc
 \includegraphics[scale=0.80]{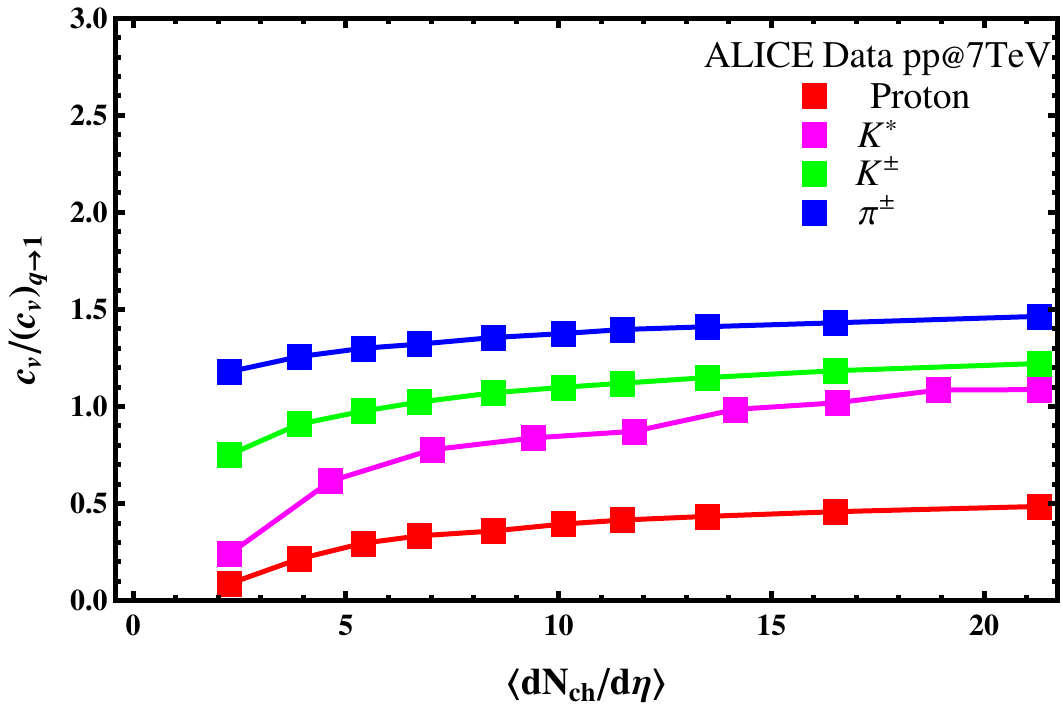}
\caption{(Color online) Heat capacity scaled by the mass density is plotted 
as a function of charged particle multiplicity (see text). The errors are within the marker size.
}
\label{fig5} 
 \ec
 \eef
\subsection{Multiplicity dependence of CSBM, speed of sound and mean transverse momentum}
  \label{mult_CSBM_c_s_2}

The speed of sound is a useful quantity which helps in characterizing
the nature of interaction in a system e.g., whether it is strongly interacting  or 
not, or how much it differs from ideal gas of massless particles.
Interaction can cause change in the effective mass of constituents, thereby, changing the 
speed of sound in the medium. CSBM gives the measure of deviation from masslessness of the 
constituents (particle mass and temperature dependence of CSBM for weakly interacting 
system is discussed in~\cite{Sarwar:2015irq}). For massless particles, $c_s^2 = 1/3$, 
however for massive particles,  $c_s^2< 1/3$. 
This is due to the fact that the massive particles do not contribute to 
the change in pressure as much as they contribute to the change 
in energy of the system. Variation of these quantities with multiplicity 
is expected to capture the change in effective interaction among constituents 
with increase in number of constituents. Also the variation of $\langle p_T \rangle$ 
of a system with the number of constituents  can capture the onset of thermalization in the system.

Variation of CSBM, $c_s^2$ and $\langle p_T \rangle$ with multiplicity for $p+p$ collisions 
have been estimated with the help of Eqs.~\ref{eq8},
~\ref{eq9},~\ref{eq10} respectively. 

It may be noted here that  $(\epsilon-3P)$ is zero for
massless ideal gas, therefore, its  non-zero value is
a measure of interaction in the system. 
Fig.~\ref{fig6} shows variation of CSBM [$\sim (\epsilon - 3P)/T^4$] 
of $\pi^{\pm}$, 
$K^{\pm}$, $K^{*0}+\overline{K^{*0}}$ and $p + \overline{p}$ with $\langle dN_{\rm ch}/d\eta \rangle$. 
It is observed that the CSBM for pions slowly reduces and as multiplicity increases while for kaons and $K^{*}$ 
it shows almost remains constant for $\langle dN_{\rm ch}/d\eta \rangle >5$ (within error bars).   
CSBM displays an increasing behaviour with $\langle dN_{\rm ch}/d\eta\rangle > 5$. 
In comparison with PYTHIA8 
generated results, we observed that $\pi^{\pm}$ and $K^{\pm}$ trend underestimates the ALICE data 
while to some extent PYTHIA8 explains $K^{*0} + \overline{K^{*0}}$ and $p+\overline{p}$. It is expected that for a thermalized medium, the contribution of a hadron of particular species to CSBM peaks when the temperature of the system is half of its mass~\cite{Sarwar:2015irq}. The value of $T$ obtained from the present analysis  is less than 190 MeV~\cite{Khuntia:2018znt}. Therefore, for pions the peak
in CSBM can be achieved for
$T\geq m_\pi/2$.
However, all other hadrons can not achieve the peak in  
CSBM as they are heavy and $T<m_H/2$, where $m_H$ is mass
of the hadrons heavier than pion. 
The larger values of CSBM indicates significant amount of interactions among hadrons or
pressure is low in the non-relativistic limit.

Fig.~\ref{fig7} shows the variation of $c_{s}^{2}$ (Eq.~\ref{eq9}) of $\pi^{\pm}$, $K^{\pm}$, $K^{*0} + 
\overline{K^{*0}}$ and $p + \overline{p}$ as a function of $\langle dN_{\rm ch}/d\eta\rangle $. 
It is observed that as we move from low to high-multiplicity of ALICE data, the $c_s^2$ for 
$\pi^{\pm}$ almost remains constant, while $c_s^2$ for $K^{\pm}$ increases upto $\langle dN_{\rm ch}/d\eta \rangle \approx$ 4 
and then saturates. $c_s^2$ for $K^{*0}+\overline{K^{*0}}$ and $p+\overline{p}$ increase with  
multiplicity. It is also observed that PYTHIA8 overestimates the ALICE data. 
As expected, low mass particles will have higher $c_{s}^{2}$ than heavier mass particles. 
The saturated value of  $c_{s}^{2}$ beyond $\langle dN_{\rm ch}/d\eta \rangle\approx 6$ follows the mass ordering. The results obtained from PYTHIA8 data, however, is less than the values obtained from experimental data.
%This indicates a possible formation of a thermalized system formed in high-multiplicity 
%$p+p$ collisions whereas PYTHIA8 contains no such medium.

Fig.~\ref{fig8} shows $\langle p_{T} \rangle$ of $\pi^{\pm}$, $K^{\pm}$,
 $K^{*0} + \overline{K^{*0}}$
and $p+\overline{p}$ as  a function of $\langle dN_{\rm ch}/d\eta \rangle$ for ALICE data and PYTHIA8 
generated results estimated using Eq.~\ref{eq10} with lower limit of integration
varying from $0.17$ to $0.22$ GeV/c to reproduce the  $\langle p_{T} \rangle$ reported 
in Ref.~\cite{Acharya:2018orn} (this limit on integration is now used for all other 
calculations for $\langle p_{T} \rangle$, which have no apparent effect on other 
observables considered here). It is observed that $\langle p_{T} \rangle$ of 
all hadrons increase very slowly  as multiplicity increases. 
%Saturation seems to set in from $\langle dN_{\rm ch}/d\eta \rangle \approx$ 5 onwards. 
The higher are the mass of hadrons, 
the higher are the values of  $\langle p_{T} \rangle$. This may be indicative of the 
presence of collectivity in the system through transverse flow as higher mass hadrons get 
affected by the flow more ($p_T\sim m v_T$
where $m$ is the mass of the hadrons and $v_T$ is the transverse flow
velocity). 

\vspace{0.05em}

\bef[ht]
 \bc
 \includegraphics[scale=0.440]{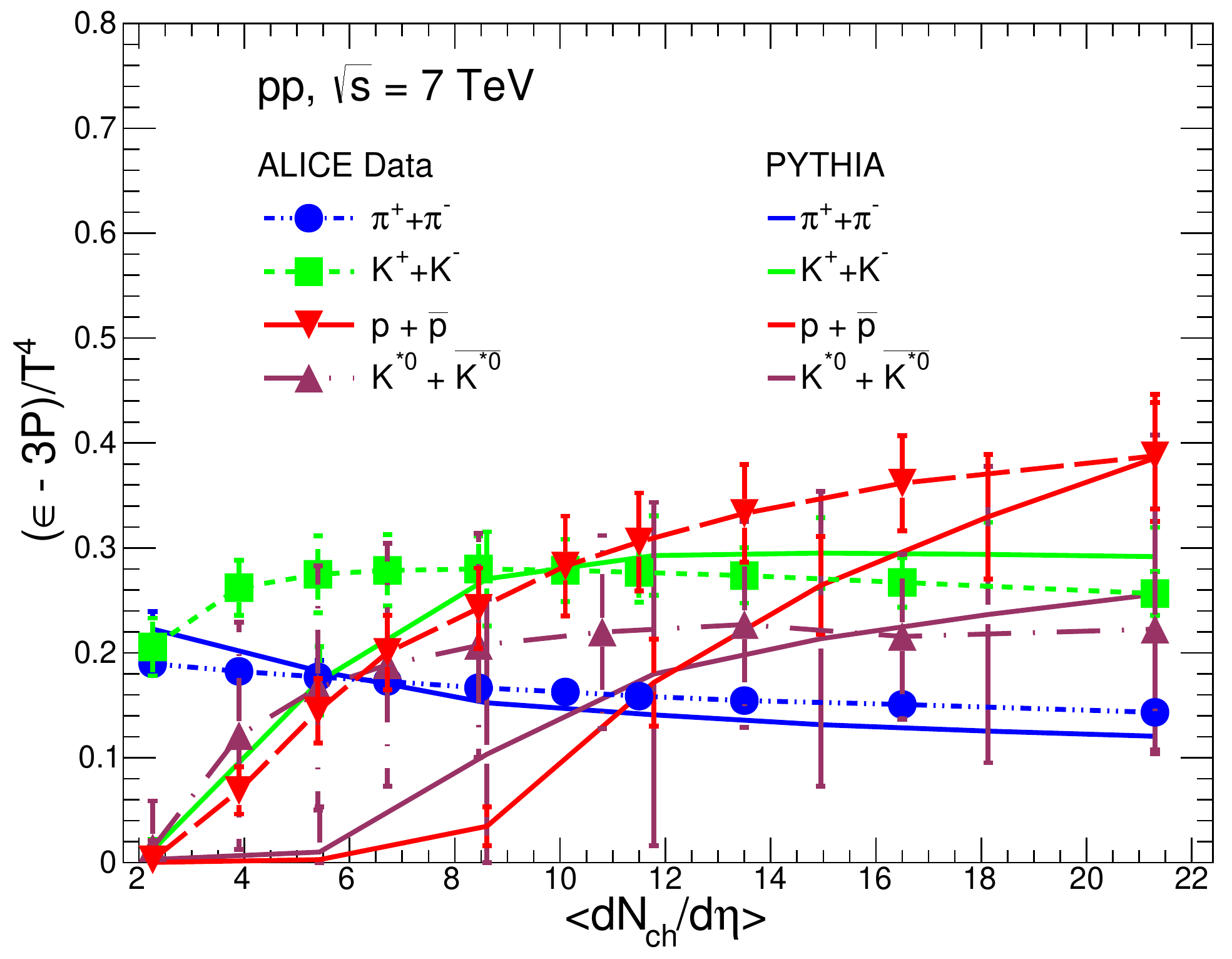}
 \caption  {(Color online) Variation of CSBM with $\langle dN_{\rm ch}/d\eta\rangle$ is shown.
Dashed (solid) lines represent results obtained by using ALICE data (PYTHIA8 simulation)
for $p+p$ collisions at $\sqrt{s}$ = 7 TeV.
}
\label{fig6} 
 \ec
 \eef

%\vspace{50.5em}

\bef[ht]
 \bc
 \includegraphics[scale=0.445]{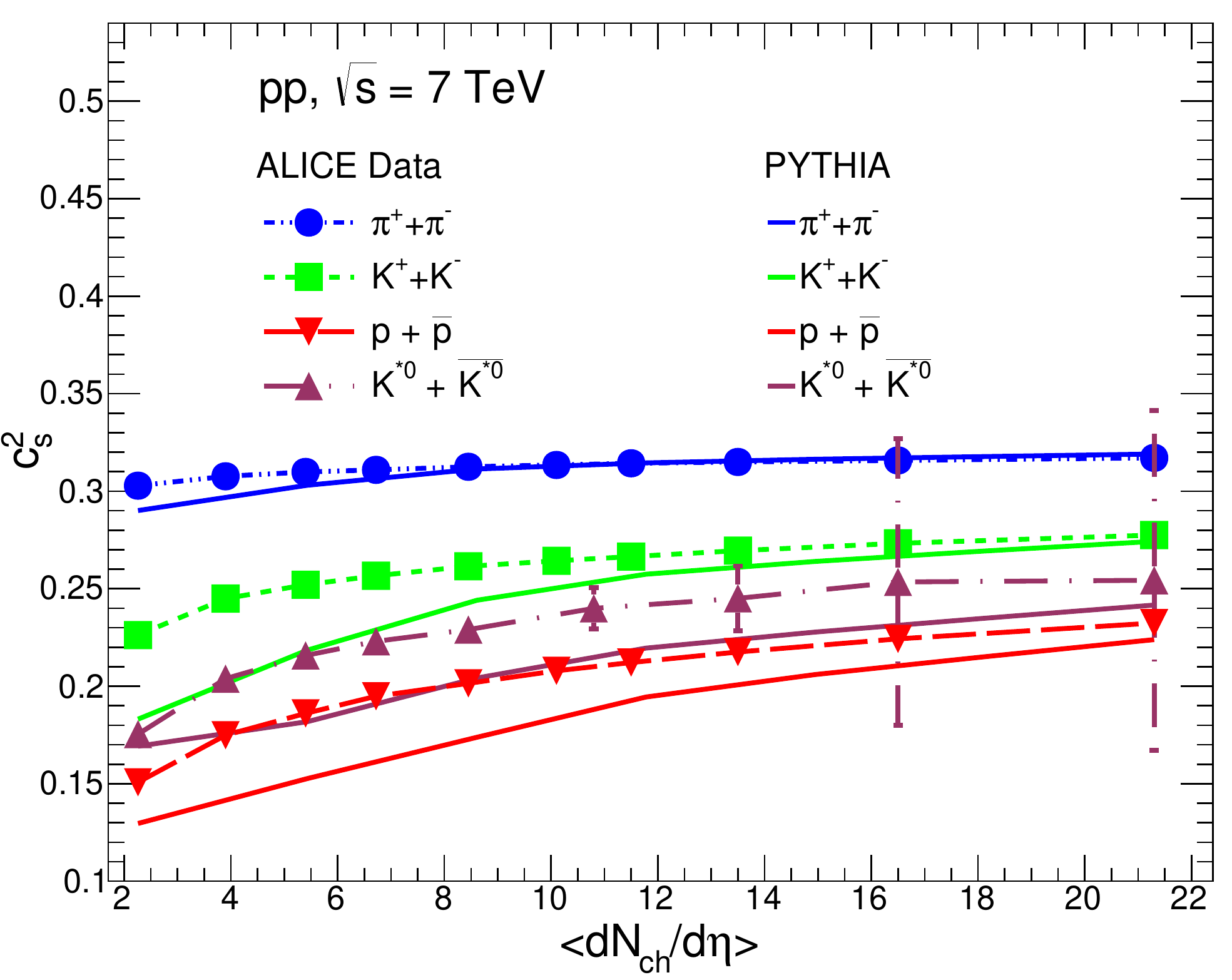}
 \caption  {(Color online) Same as Fig.~\ref{fig6} for speed of sound.}
\label{fig7} 
 \ec
 \eef

\bef[ht]
 \bc
 \includegraphics[scale=0.445]{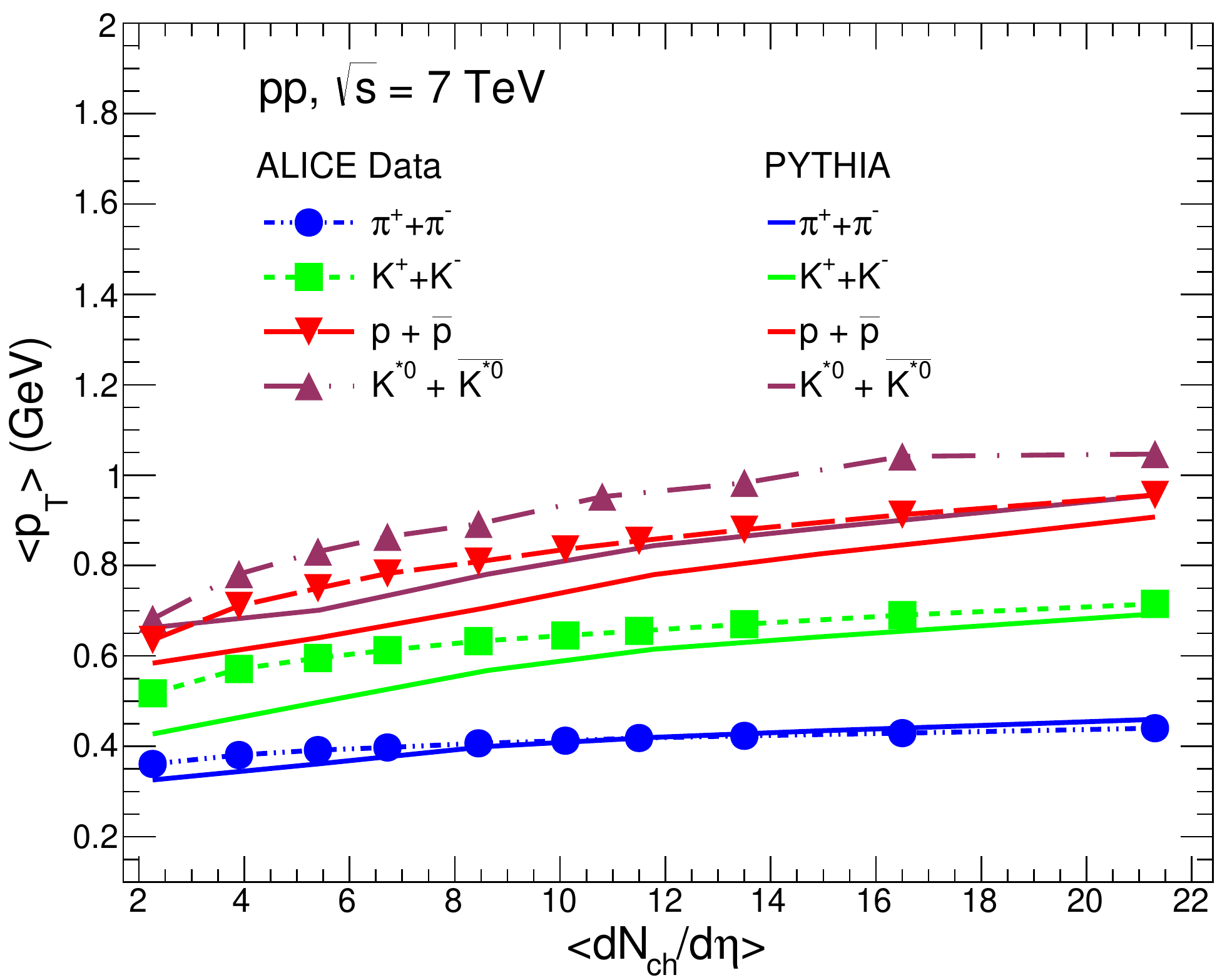}
 \caption  {(Color online) Mean transverse momentum in GeV is presented as a function 
of charged particle multiplicity. Dashed (solid) lines are obtained by 
using ALICE data (PYTHIA8 simulation).} 
\label{fig8} 
 \ec
 \eef

 It is interesting to find that all of the above quantities 
for lighter hadrons show saturation for
$\langle dN_{\rm ch}/d\eta\rangle$ $\geq$ (4-6) 
in their variation with 
$\langle dN_{\rm ch}/d\eta\rangle$. This general feature may be the hint of onset of 
their possible randomized collective nature.
This is more prominent in the variation of speed of sound and CSBM with
multiplicity, where heavier hadrons show different trends from that of lighter 
hadrons. Moreover, the
saturation found here is vastly different from the saturation of $\langle p_{T} \rangle$ of all
charged particles which occurs at $\langle dN_{\rm ch}/d\eta \rangle$ $\approx$ 20 as in Ref.~\cite{chargmpt}. This may be due to the inclusion of heavier
particles in calculation of $\langle p_{T} \rangle$. In fact, in this work, it is found that heavier particles like proton
shows different nature; for them instead of saturation, quantities considered here increases
monotonically.
%due to their separation from the
%produced system in the early stage. They do not show saturation, except for proton the slope
%reduces at $N_{ch}$ $\approx$ (4-6) for $c^{2}_{s}$ and CSBM. 
It is also interesting to note that heavier hadrons
are described well by PYTHIA8.  
It further emboldens the possibility of formation of strongly correlated but
randomized medium, as this can not be explained by color reconnection (CR) effect of final state which is included in PYTHIA8. This mismatch points out something more than CR effect is responsible for such saturation, hinting scope of presence of collectivity in the system from which these particles originate.

 %\vspace{0.005em}
 %-----------------------------------------------------finite system size effect on Heat capacity-----------------------------------------------------------------------------------------------------------

\subsection{Finite system size dependence of heat capacity}
\label{finite_size_effect}
In case of RHICE, the thermal nature of produced particles is extensive type (BG), but for 
$p+p$ collisions, Tsallis (TB) distribution fits the particle spectra very well~\cite{Marques:2015mwa,Cleymans:2012ac}. The appearance 
of non-extensive statistics in a system may be for several reasons e.g., finite size effect, 
long-range interaction or
correlation. For this reason, in this work, 
it is investigated whether finite size effect 
alone can explain the 
deviation of the value of $q$ from unity.
We incorporate the finite-size effect by considering a lower momentum cutoff, 
$p_{min}=\pi/R$, in the momentum integration, where, R is the radius of the 
system~\cite{Bhattacharyya:2015zka}. As the collision energy is the same, large multiplicity events 
are expected to be originating from larger overlap region in $p+p$ collisions. 
We have considered different radius (R) with each multiplicity following the relation R $\sim \langle dN_{\rm ch}/d\eta \rangle^{1/3}$ \cite{Aamodt:2011mr,Abelev:2014pja,Adam:2016bpr}. 
The $R$ dependence of $C_V$, CSBM, $c_s$ and $\langle p_T \rangle$ have been extracted 
by fitting data from $p+p$ collision at $\sqrt{s}=7$ TeV to TB distribution
with $T$ and $q$ as fitting parameters. The data sets have also been studied by
using BG statistics (in the limit $q\rightarrow 1$) 
with the same value of $T$ obtained from TB statistics,
to check whether extensive TB distribution with finite size effect can account 
for the q-value extracted by fitting experimental data.
%We compare values of the observables at each multiplicity, calculated using TB 
%distribution with T and  q values extracted from ALICE data, with values calculated 
%using Boltzmann's distribution considering same temperature but with $q=1$, where, 
%while doing the momentum integration, the finite size is taken into account. 

In order to account for the effects of system size, we have studied variation of heat capacity, 
heat capacity scaled by average number of particles and  $T^3$ with finite system size using Eq.~\ref{eq7}. We find the lower limit of $R$ as 1.3 fm and the upper limit to be 2.7 fm. This is used to represent the available multiplicity classes such that the values with $q \neq 1$ same as that of earlier plots showing variation with multiplicity. Finite system size is also reflected through the value of $q>1$ in contrast to $q\rightarrow 1$.

Fig.~\ref{fig9} shows $C_{V}$ of $\pi^{\pm}$, $K^{\pm}$, $K^{*0} + \overline{K^{*0}}$ 
and $p + \overline{p}$ obtained by using ALICE data as a function of system size. 
It is observed that  the $C_{V}$ of $\pi^{\pm}$, $K^{\pm}$ and $p+\overline{p}$ 
increases with system size for $q\neq 1$. 
The slope of $C_{V}$ for $\pi^{\pm}$ is less compared to $K^{\pm}$ and $p+\overline{p}$. Results with $q\rightarrow 1$ (corresponding to BG statistics) represented by solid curves indicate that $C_{V}$ of $K^{\pm}$, $p+\overline{p}$ are underestimated  by PYTHIA8 unlike $\pi^{\pm}$. 

Fig.~\ref{fig10} shows $C_{V}$ scaled by $T^{3}$ for $\pi^{\pm}$, $K^{\pm}$, $K^{*0}+
\overline{K^{*0}}$ and $p+\overline{p}$ extracted from ALICE data as a function of system size. 
It is observed that $C_V/T^3$ for $\pi^{\pm}$, $K^{\pm}$, $K^{*0}+\overline{K^{*0}}$ and $p+\overline{p}$ 
vary slowly  with increasing system size for both with TB and BG statistics (except for $p+\overline{p}$).  

\bef[ht]
\bc
\includegraphics[scale=0.445]{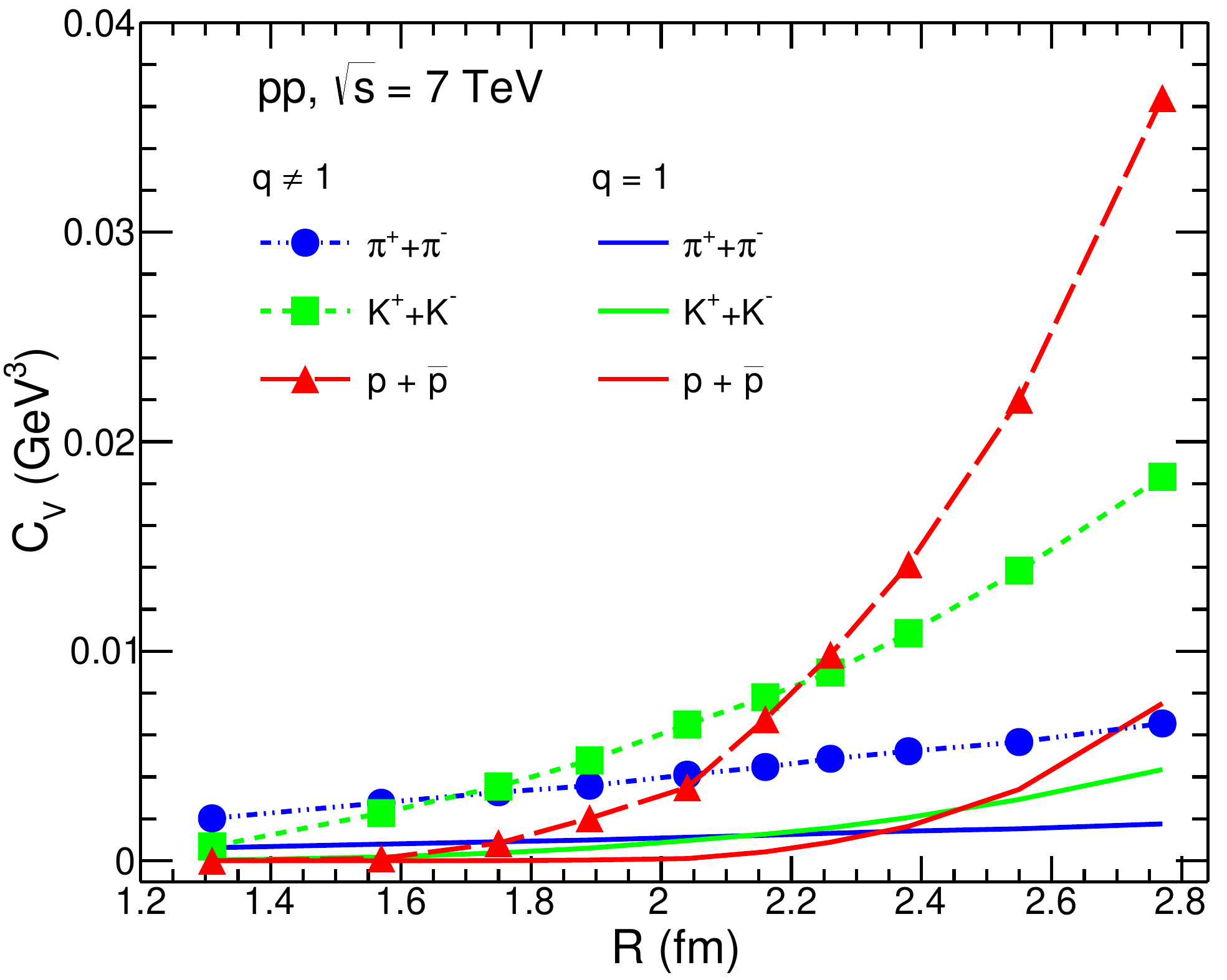}
% CvvR.pdf}
\caption {(Color online) Heat capacity obtained by using BG and TB distribution as a function of 
system size. Dashed (solid) lines represent results for $q \neq 1$ ($q=1$)for 
$p+p$ collisions at $\sqrt{s}$ = 7 TeV.}
 \label{fig9}  
 \ec
 \eef

\bef[ht]
\bc
\includegraphics[scale=0.445]{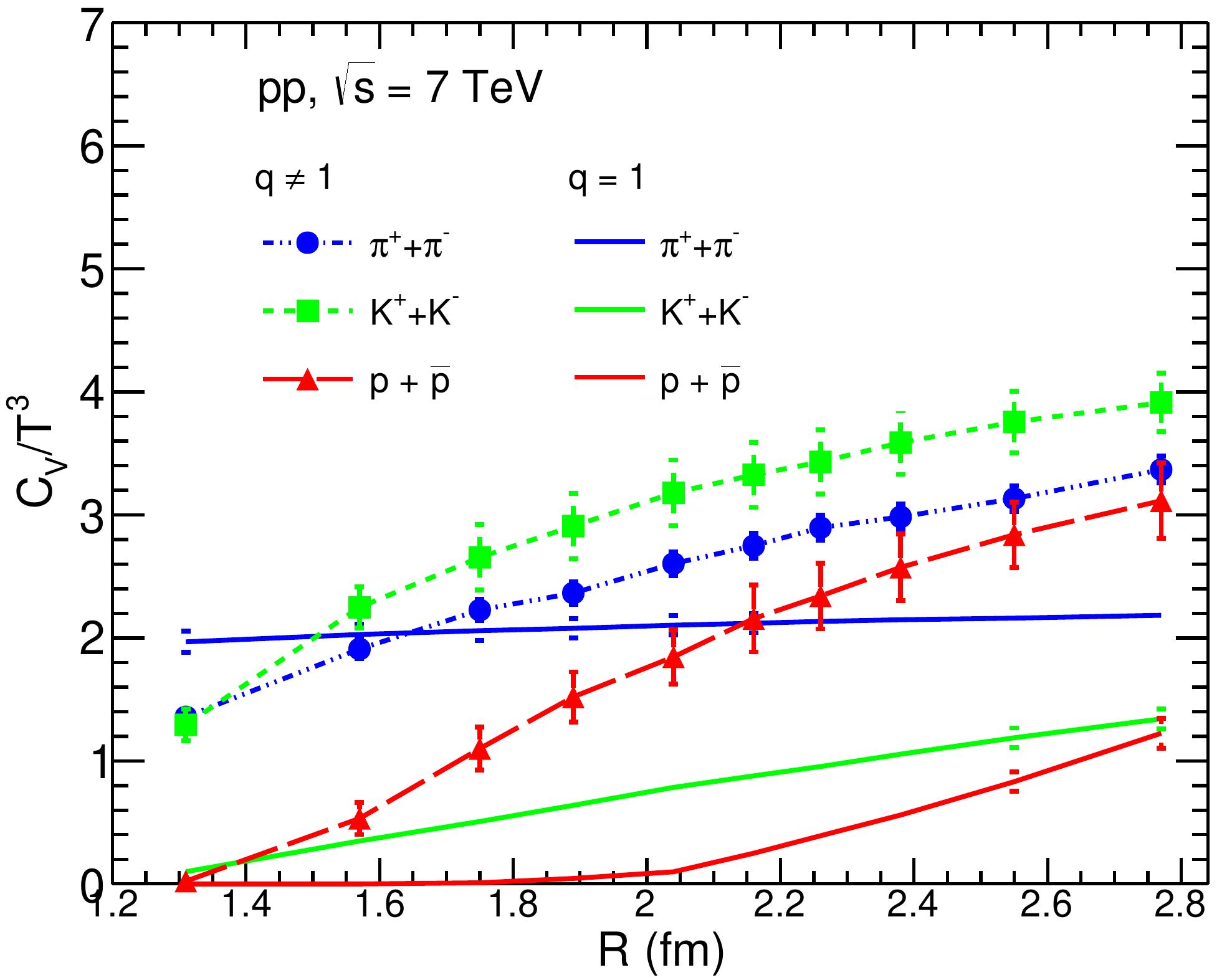}
%CvbT3vR.pdf}
\caption {(Color online) Same as Fig.~\ref{fig9} for heat capacity scaled by $T^3$.
}
\label{fig10}  
 \ec
 \eef      

  %-----------------------------------------------------finite system size effect on CSB, speed of sound, mean pT-------------------------------------------------------------------------------   

\subsection{Finite system size dependence of CSBM, speed of sound and mean transverse momentum}

Fig.~\ref{fig11} shows CSBM of $\pi^{\pm}$, $K^{\pm}$ and $p+\overline{p}$ obtained from ALICE 
data as a function of system size. Results with $q\neq 1$ represented by dashed lines, it is observed 
that CSBM of $\pi^{\pm}$ decreases slowly while CSBM of $K^{\pm}$ increases slightly at 
small $R$. But $p+\overline{p}$ displays an 
increasing trend.
Results for BG statistics represented by solid curves show similar trend.
It may be noted that 
the heavier hadrons contribute more to the energy density than pressure through their rest mass energy, 
therefore for proton $(\epsilon-3P)$ will be more than pions.  
%The slope of CSBM of $\pi^{\pm}$  increases almost linearly with R. 

Fig.~\ref{fig12} shows  $c_{s}^{2}$ for $\pi^{\pm}$, $K^{\pm}$, $K^{*0}+\overline{K^{*0}}$ and $p+\overline{p}$  
extracted from  ALICE data as a function of system size. The $c_s^2$ shows a plateau as a function of $R$ both 
for BG and TB statistics for all the hadronic species.  

Fig.~\ref{fig13} shows $\langle p_{T} \rangle$ for $\pi^{\pm}$, $K^{\pm}$, $K^{*0}+\overline{K^{*0}}$ and $p+\overline{p}$ of ALICE data as a function of system size. For TB statistics, it is observed that $\langle p_{T} \rangle$ of all hadrons increases with system size (faster for $p+\overline{p}$) and  reaches a plateau beyond $R\sim 1.8$ fm.  In BG statistics, $\pi^{\pm}$ and $K^{\pm}$ show slow
variation. 

It is generally observed that the incorporation of finite size effect in BG statistical approach 
can not reproduce the value of the observables calculated with non-extensivity parameter ($q$) 
extracted from the $p+p$ collisions. This may suggest that the appearance of non-extensivity in 
$p+p$ collisions may not be completely explained by finite size effect alone, 
thereby hinting the presence of other physical effects like long-range correlation that also 
contributes to the origin of non-extensivity.

\bef[ht]
\bc
\includegraphics[scale=0.445]{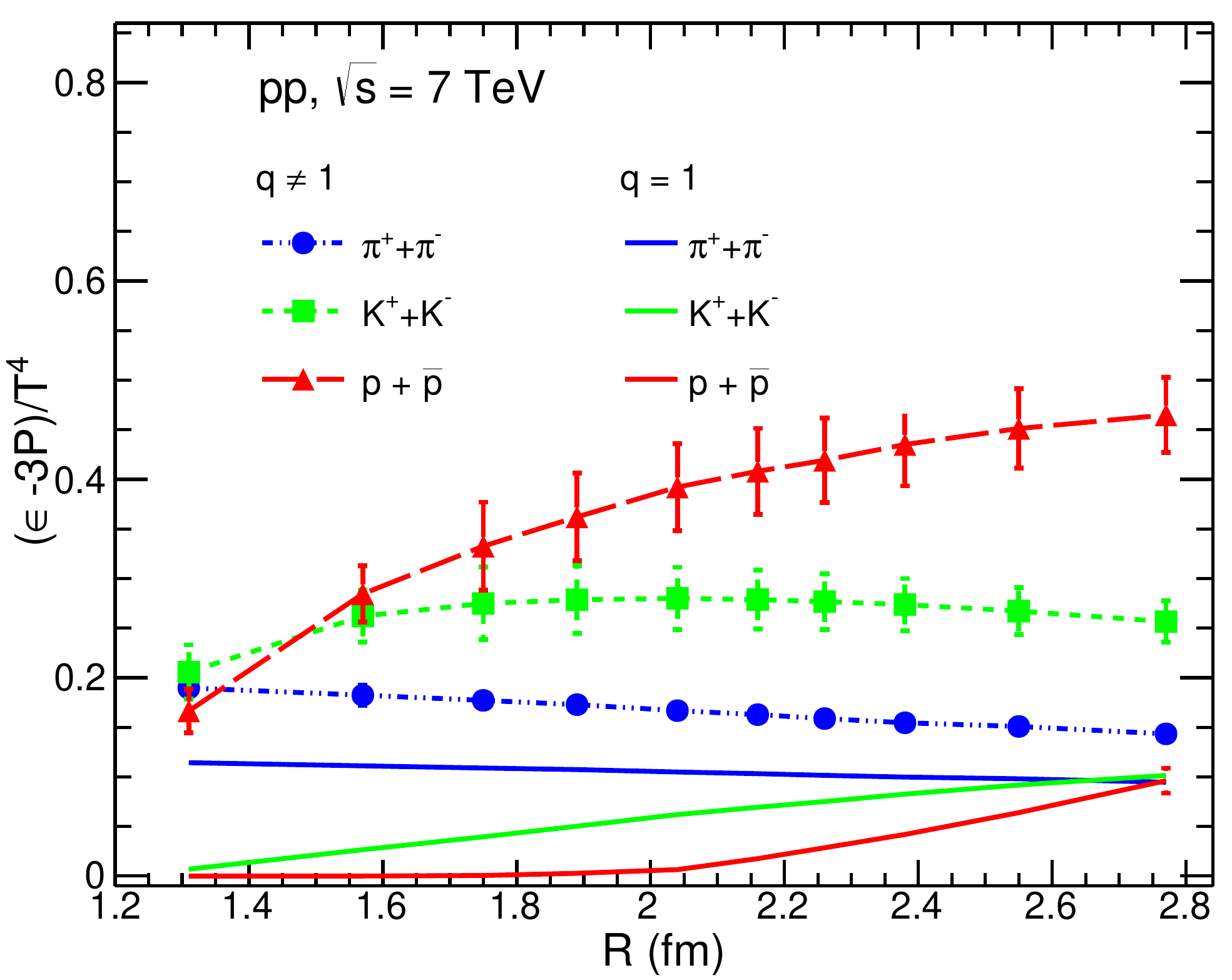}
%CSBvR.pdf}
\caption {(Color online) Same as Fig.~\ref{fig9} showing variation of  CSBM with $R$.
}
 \label{fig11}  
 \ec
 \eef

\bef[ht]
\bc
\includegraphics[scale=0.445]{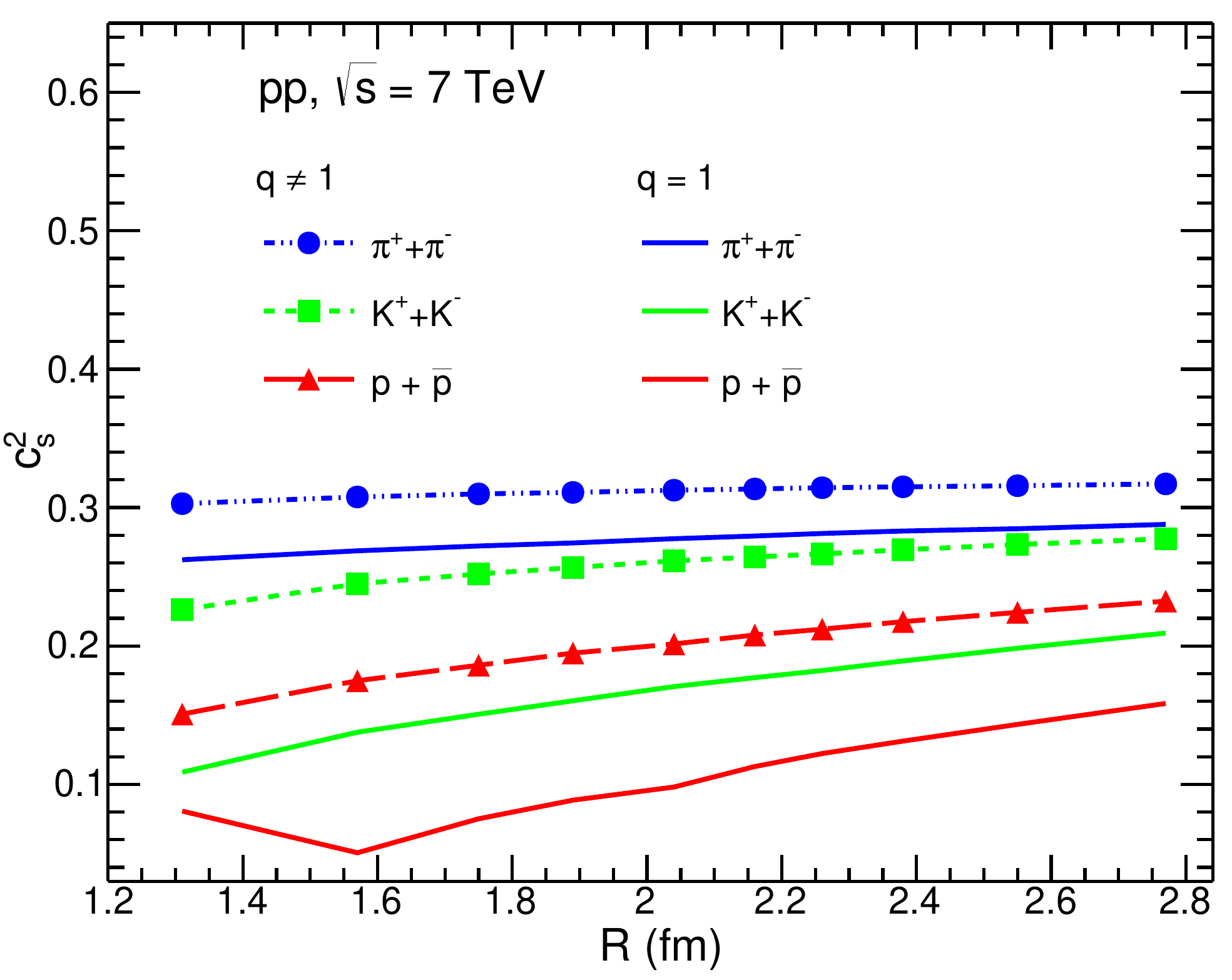}
%Cs2vR.pdf}
\caption {(Color online) Same as Fig.~\ref{fig9} for $c_s^2$.
}
 \label{fig12}  
 \ec
 \eef

\bef[ht]
\bc
\includegraphics[scale=0.445]{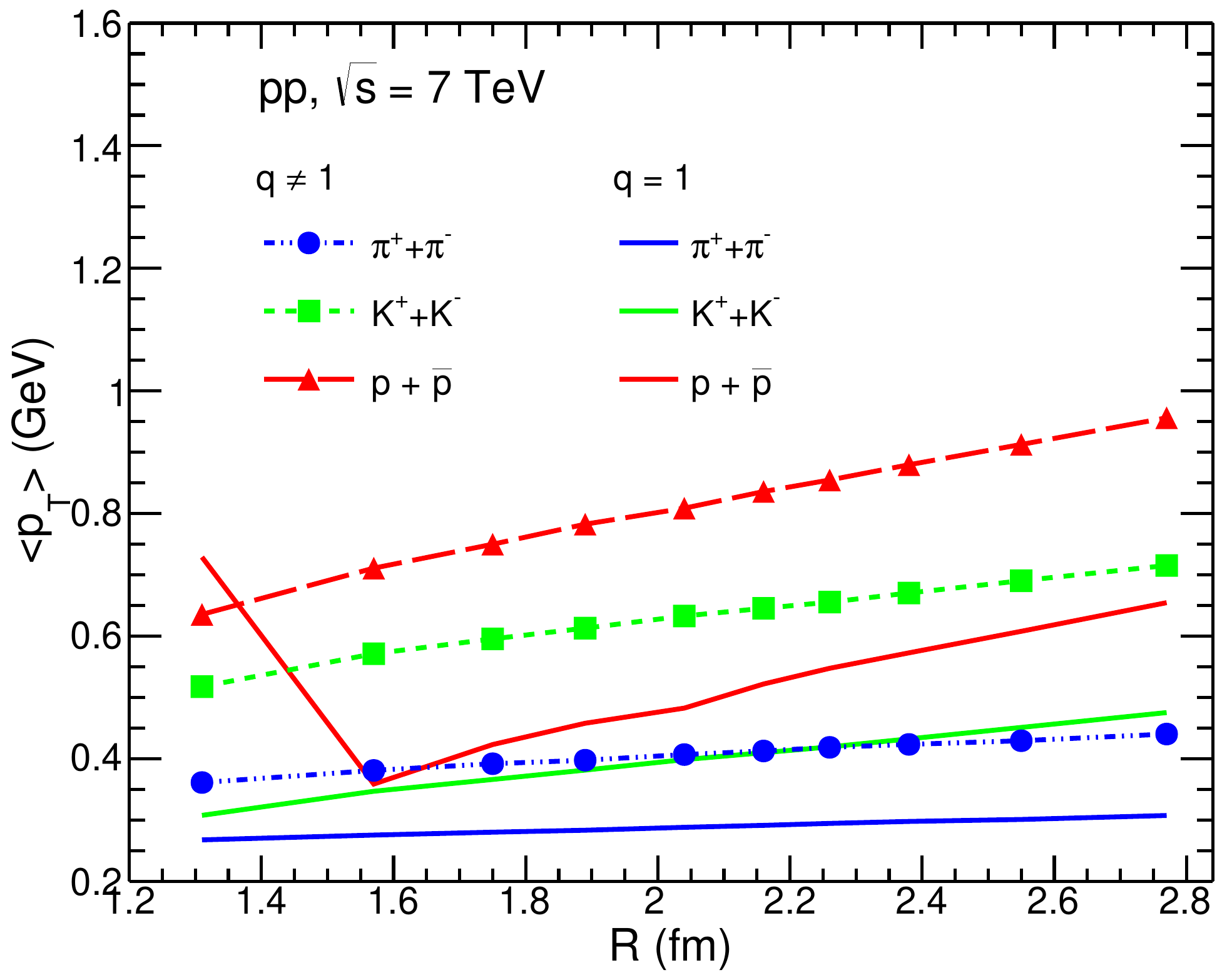}
%MptvR.pdf}
\caption {(Color online) Same as Fig.~\ref{fig9} for $\langle p_T\rangle$.
}
 \label{fig13}  
 \ec
 \eef

 %-----------------------------------------------------centre of mass energy dependence on heat capacity-------------------------------------------------------

 \subsection{Energy dependence of $C_{V}$, $CSBM$, $c_s^2$}
\label{enrg_Cv}
The collision energy dependence of heat capacity scaled by average number of particles and $T^3$ obtained from RHICE and ALICE $p+p$ data at different $\sqrt{s}$ has
been studied by using the values of $T$ and $q$ extracted from the TB distribution fit of the $p_{T}$-spectra~\cite{Saraswat:2017kpg}, where $\sqrt{s}$ ranges from 0.0624 TeV to 13 TeV. 

Fig.~\ref{fig14} shows $C_{V}$ scaled by the average density of charged 
pions ($\langle n_{\pi} \rangle= \langle n_{\pi^+}\rangle+ \langle n_{\pi^-} \rangle$) as a function of $\sqrt{s}$. 
Here, scaling by $\langle n_{\pi} \rangle$ is considered as 
production of ($\pi^{+}$ and $\pi^{-}$) is abundant in relativistic collisions. 
It is observed that $C_{V}/\langle n_{\pi} \rangle$ increases sharply upto $\sqrt{s}=1.5$ TeV beyond which it 
increases very slowly. 
In the same figure we also display the variations of 
$C_{V}/ T^{3}$ and  CSBM  of charged particles obtained from RHICE and ALICE data as a function of $\sqrt{s}$. 
Both the quantities tend to saturate (within error bars) for $\sqrt{s}>2$ TeV. 
We find that speed of sound seems to be almost constant (Fig.~\ref{fig14})  for
$\sqrt{s}>2$ GeV.  
Possibly for $p+p$ collisions with $\sqrt{s}\leq 2$ TeV 
a thermal medium is formed with  
the value of $c_s^2\approx 0.24$. Such a value of $c_s^2$ is obtained  
in hadronic resonance gas model calculation~\cite{Sarwar:2015irq}.

The general observation in this regard is that the thermodynamic quantities considered here 
show a saturation starting for	 $\sqrt{s}\geq 2$ TeV. The nature of variation of $C_V/\langle n_\pi\rangle$
beyond $\sqrt{s} \approx 2$ TeV is similar to that found in heavy-ion collisions at the 
chemical-freeze out surface as in Ref.~\cite{Basu:2016ibk}. Therefore, this may be taken
as a  hint for the formation of medium similar in kind to that of heavy-ion collisions.
Indicating that for $\sqrt{s}\geq $ 2 TeV sufficient number of particles are
produced to form  QCD medium. It is interesting to further note that the average multiplicity 
for $\sqrt{s} \approx 1.5$ lies between 3 to 7 as in Ref.~\cite{Alkin:2017zob}. This again puts 
weight to the possibility that the saturation effect as observed in variation of above 
thermodynamic quantities with multiplicity is potentially due to formation of a medium in 
$p+p$ collisions for multiplicity, $\langle dN_{\rm ch}/d\eta \rangle \ge 4-6$. We note that for observing saturation effects 
in PYTHIA8 simulated results (in which CR is thought to be responsible for the saturation), 
this kind of  saturation starts at $\langle dN_{\rm ch}/d\eta \rangle \approx 20$.   
%The particle spectra used for studying the variation for $p+p$ collisions do not have any 
%information regarding, which freeze out surface they have originated from. 
%However, the trend matches with that of heavy-ion collisions for its chemical 
%freeze out surface. This fact may be due to vanishing time difference for occurrence of 
%chemical and kinetic freeze out surface in $p+p$ collisions.

 \bef[ht]
 \bc
  \includegraphics[scale=0.445]{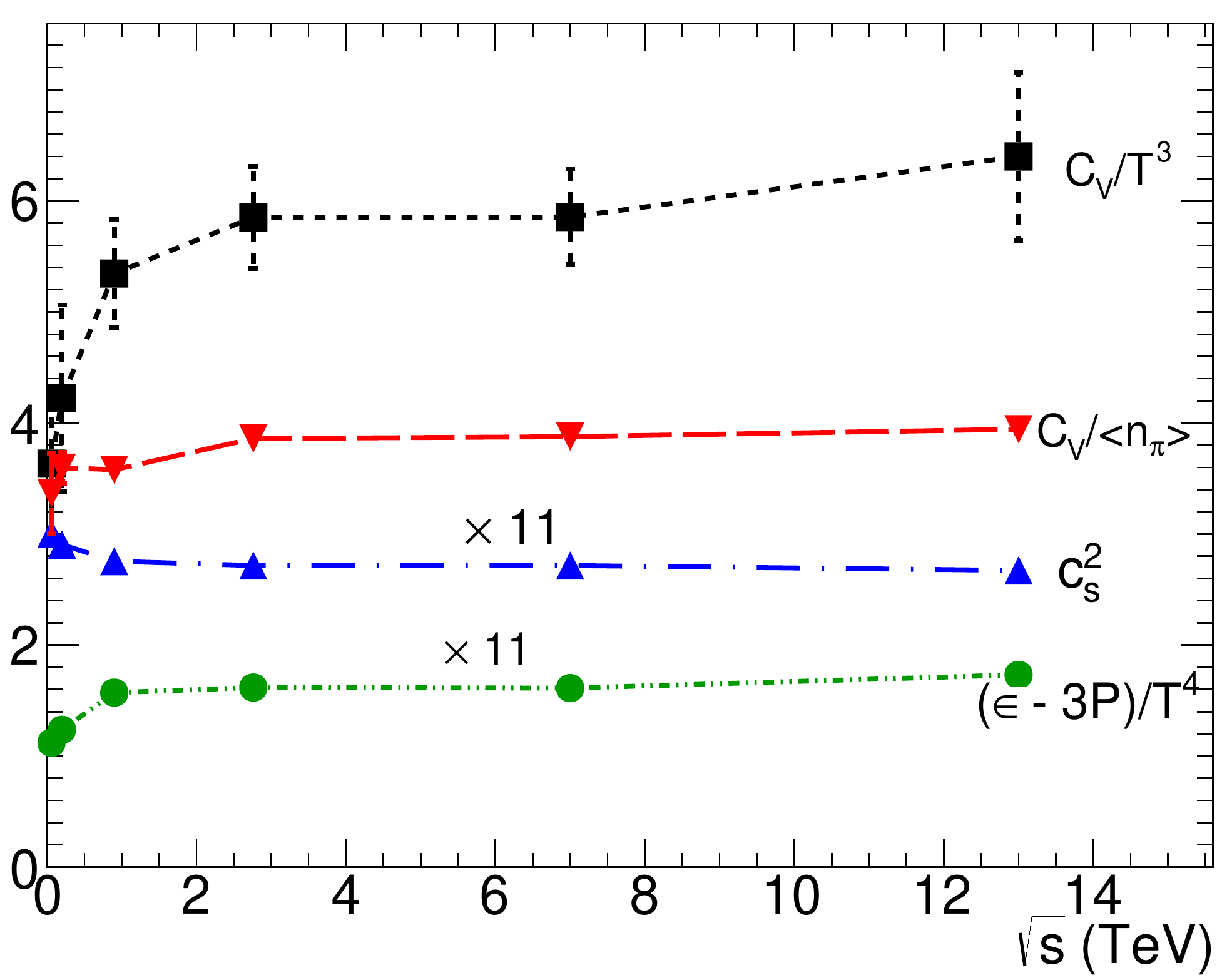}
\caption{(Color online) Variation of heat capacity scaled by $T^3$ and average charged pion density ($\langle n_{\pi} \rangle$), speed of sound and CSBM with $\sqrt{s}$ for $p+p$ collisions.}
 \label{fig14}  
 \ec
 \eef 

 \iffalse
\bef[ht]
\bc
 \includegraphics[scale=0.95]{CvbyT3withRootS.pdf}
 \caption{(Color online) Same as  Fig.~\ref{fig14} for $C_V/T^3$.
 }
 \label{fig15}
 \ec
 \eef

\bef[ht]
\bc
\includegraphics[scale=0.89]{CSBMwithRootS.pdf}
 \caption{(Color online) Same as  Fig.~\ref{fig14} for CSBM.
 }
 \label{fig16}  
 \ec
 \eef

\bef[ht]
\bc
\includegraphics[scale=0.89]{Cs2withRootS.pdf}
 \caption{(Color online) Same as  Fig.~\ref{fig14} for $c_s^2$.
 }
 \label{fig17}  
 \ec
 \eef   
\fi
%\vspace{20.5em}
%     \vspace{50.5em}

\section{Summary}
  \label{sum}
The findings of the present work based on the analysis of data from $p+p$ collision at LHC energies may be summarized as follows:

\begin{itemize}

\item[$\bullet$]
We have analyzed how a system produced in $p+p$ collision 
at relativistic energies evolves into a collective medium as the
the number of produced particles and collision energy increases. 
For the purpose of this analysis, the thermodynamic quantities like $C_V$,
$c_s$ and CSBM have been chosen for reasons explained in the text. 
We observe that  $C_V$ achieves a plateau for $\langle dN_{\rm ch}/d\eta \rangle > (4-6)$.
We also note that $C_V/\langle n_i \rangle$ for pionic and kaonic matter have similar values. 

\item[$\bullet$ ] We have also investigated how conformal symmetry breaking/trace anomaly varies with the degrees of freedom in an  environment of QCD many body system. Similar to $C_{V}$, a saturation in CSBM with $\langle dN_{\rm ch}/d\eta \rangle$ and $\sqrt{s}\geq $ 2 TeV is also observed. 

\item[$\bullet$] The importance of high-multiplicity ($\langle dN_{\rm ch}/d\eta\rangle > (4-6)$) for medium formation in small systems is further endorsed by the observation of similar kind of saturating behaviour of the thermodynamic quantities considered here with collision energies. This suggests that at collision energies, $\sqrt{s}\geq $ 2 TeV,  the sea quarks and gluons within the proton are large enough to produce QCD  medium.
%higher collision energies the sea quarks and gluons within the proton are large enough to produce QCD  medium.

 \item[$\bullet$] Comparisons of the results extracted from ALICE data with the results obtained from PYTHIA8 simulation have been carried out. It is observed that PYTHIA8 (devoid of medium) explains scaled $C_{V}$ for heavy particles approximately but  it cannot explain the trend of lighter hadrons. This may be a sign that lower mass particles originate  from a thermalized  medium. 

\item[$\bullet$] The deviation in the value of $q$ from unity in TB statistics may indicate the presence of long-range correlations as well as the finiteness of the system. However, it is observed that finite size effect alone cannot account for the appearance of $q \neq 1$ value. This may suggest that the presence of effects other than finiteness {\it  e.g.}, correlations,  in QCD system play important role for giving rise to non-extensivity.

 \end{itemize}

\section{Acknowledgement} 
 SD, GS and RNS acknowledge the financial supports  from  ALICE  Project  No. SR/MF/PS-01/2014-IITI(G) of Department  of  Science $\&$ Technology,  Government of India. Further, R.S. acknowledges the financial supports from DAE-BRNS Project No. 58/14/29/2019-BRNS. JA is grateful to Tramabak Bhattacharyya for useful discussions.

\vspace{10.005em}


\begin{thebibliography}{}

%\bibitem{hatsuda} K. Yagi, T. Hatsuda and Y. Miake, Quark Gluon Plasma: 
%From Big Bang to Little Bang, Cambridge Monographs on Particle Physics, Nuclear Physics
%and Cosmology, 2008.

\bibitem{Baier:2006um} 
   R.~Baier, P.~Romatschke and U.~A.~Wiedemann,
  %``Dissipative hydrodynamics and heavy ion collisions,''
  Phys.\ Rev.\ C {\bf 73}, 064903 (2006).

  \bibitem{Alver:2010rt} 
  B.~Alver {\it et al.} [PHOBOS Collaboration],
  %``Non-flow correlations and elliptic flow fluctuations in gold-gold collisions at $\sqrt{s_{NN}}=200$ GeV,'' 
  Phys.\ Rev.\ C {\bf 81}, 034915 (2010).

 \bibitem{Kopeliovich:2011zz} 
  B.~Z.~Kopeliovich, I.~K.~Potashnikova and I.~Schmidt,
  %``Nuclear suppression of J/Psi: from RHIC to the LHC,''
  Nucl.\ Phys.\ A {\bf 864}, 203 (2011).

 \bibitem{Agakishiev:2011ar} 
  G.~Agakishiev {\it et al.} [STAR Collaboration],
  %``Strangeness Enhancement in Cu+Cu and Au+Au Collisions at $\sqrt{s_{NN}} = 200$ GeV,''
  Phys.\ Rev.\ Lett.\  {\bf 108}, 072301 (2012).

 \bibitem{Isobe:2006vg} 
  T.~Isobe [PHENIX Collaboration],
  %``Measurement of high-p(T) hadrons at RHIC-PHENIX,''
  nucl-ex/0605016.

\bibitem{Ridge1}  V. Khachatryan, et al., CMS, J. High Energy Phys. {\bf 09}, 091 (2010).

\bibitem{Ridge2} W. Li, Mod. Phys. Lett. A {\bf 27}, 1230018 (2012).
\bibitem{Ridge3} V. Khachatryan, et al., CMS, Phys. Rev. Lett. {\bf 116}, 172302 (2016).

\bibitem{Ridge4} G. Aad, et al., ATLAS, Phys. Rev. Lett. {\bf 116}, 172301 (2016).

\bibitem{Ridge5}V. Khachatryan, et al., CMS, Phys. Lett. B {\bf 765}, 193 (2017).

\bibitem{Zhao:2017rgg} W.~Zhao, Y.~Zhou, H.~Xu, W.~Deng and H.~Song, Phys.\ Lett.\ B {\bf 780}, 495 (2018).

  %``Hydrodynamic collectivity in proton+proton collisions at 13 TeV,''

 \bibitem{MarkMac:2019plb} M. Mace, V.V. Skokov, P. Tribedy, R. Venugopalan, Phys. Rev. Lett. {\bf 121}, 052301 (2018).

\bibitem{Braun:2015eoa} M.~A.~Braun, J.~Dias de Deus, A.~S.~Hirsch, C.~Pajares, R.~P.~Scharenberg and B.~K.~Srivastava,
%``De-Confinement and Clustering of Color Sources in Nuclear Collisions,''
Phys. Rept. \textbf{599}, 1 (2015).

\bibitem{Landau:1965cpl} L. D. Landau, Izv. Akad. Nauk. SSSR {\bf 17}, 51 (1953);  S. Belenkij and L. D. Landau, 
Usp. Fiz. Nauk. {\bf 56}, 309 (1955):  Nuovo Cimento Suppl. {\bf 3}, 15 (1956);
D. ter Haar (Ed.), Collected papers of L.D. Landau, Gordon \& Breach, New York, 1965, p. 665.

  \bibitem{VanHove:1982vk} 
  L.~Van Hove,
  %``Multiplicity Dependence of p(T) Spectrum as a Possible Signal for a Phase Transition in Hadronic Collisions,''
  Phys.\ Lett.\  {\bf 118B}, 138 (1982).

  \bibitem{Hirsch:2018pqm}
R.~P.~Scharenberg, B.~K.~Srivastava and C.~Pajares,
%``Exploring the initial stage of high multiplicity proton-proton collisions by determining the initial temperature of the quark-gluon plasma,''
Phys. Rev. D \textbf{100}, 114040 (2019).

\bibitem{tracanomaly1}  
S. Borsanyi et al., JHEP {\bf 1011}, 077 (2010).

\bibitem{traceanomaly2}  
S. Borsanyi et al., J. Phys: Conf. Series {\bf 316},
012020 (2011). 

 \bibitem{Thakur:2019qau} 
  D.~Thakur [ALICE Collaboration],
  %``Quarkonium production as a function of charged particles multiplicity in pp and p-Pb collisions measured by ALICE at the LHC,''
  PoS HardProbes 2018, {\bf 164} (2019). 

\bibitem{jpsiSat:2018} D. Adamová et al. [ALICE Collaboration], Phys. Lett. B {\bf 776}, 91 (2018). 

\bibitem{chargmpt} B. Abelev {\it et al.} for 
ALICE Collaboration,  Physics Letters B {\bf 727}  371 (2013).

\bibitem{Nature}   J. Adams et al.  Nature Physics {\bf 13}, 535 (2017).

  
\bibitem{Basu:2016ibk} S.~Basu, S.~Chatterjee, R.~Chatterjee, T.~K.~Nayak and B.~K.~Nandi, Phys.\ Rev.\ C {\bf 94}, 044901 (2016).
  %``heat of Matter Formed in Relativistic Nuclear Collisions,''  

  
 \bibitem{Li:2007ai}  X.~M.~Li, S.~Y.~Hu, J.~Feng, S.~P.~Li, B.~H.~Sa and D.~M.~Zhou,  Int.\ J.\ Mod.\ Phys.\ E {\bf 16}, 1906 (2007).
  %``Heat capacity relevant to QGP phase transition at s(NN)**(1/2) = 200-GeV,''
  
%  \bibitem{review} W.~Busza, K.~Rajagopal and W.~van der Schee,
%%``Heavy Ion Collisions: The Big Picture, and the Big Questions,''
%Ann. Rev. Nucl. Part. Sci. \textbf{68}, 339 (2018).

\bibitem{Reif} Reif, F., Fundamentals of Statistical and Thermal Physics, Mcgraw-Hill International
Editions, Singapore, 1985.

\bibitem{KolbTurner} E. W. Kolb and M. S. Turner, The Early Universe, Addison-Wesley Publishing Co., 
Singapore, 1989.

  \bibitem{Cleymans:2014woa}
J.~Cleymans,
%``The Tsallis distribution at the LHC: Phenomenology,''
AIP Conf. Proc. \textbf{1625}, 31 (2015).

  \bibitem{Tsallis:1987eu} 
  C.~Tsallis,
  %``Possible Generalization of Boltzmann-Gibbs Statistics,''
  J.\ Statist.\ Phys.\  {\bf 52}, 479 (1988).

\bibitem{Tsallis:2008mc} 
  C.~Tsallis,
  %``Nonadditive entropy: The Concept and its use,''
  Eur.\ Phys.\ J.\ A {\bf 40}, 257 (2009).

\bibitem{Tsallis:2009zex} 
  C.~Tsallis, Introduction to Nonextensive Statistical Mechanics (Springer, 2009).

%\bibitem{CT2} M. Gell-Mann and C. Tsallis (eds.), Nonextensive Entropy - 
%Interdisciplinary Applications, Oxford University Press (2003).

\bibitem{wilk1} G. Wilk, and Z. W{\l}odarczyk, ~ Phys. Rev. Lett. {\bf 84}, 2770 (2000) .

\bibitem{wilk2} G. Wilk, and Z. W{\l}odarczyk, ~ Phys. Rev. C {\bf 79}, 054903 (2009).

\bibitem{wilk3} G. Wilk, and Z. W{\l}odarczyk, Chaos Solitons Fractals {\bf 13}, 581 (2001).

%\bibitem{ts0} T. Bhattacharyya, J. Cleymans, and S. Mogliacci, Phys. Rev. D
%{\bf 94}, 094026 (2016).
%
%\bibitem{ts1} C. Tsallis and D.J. Bukman, Phys. Rev. E {\bf 54}, R2197(R) (1996).
%
%\bibitem{ts2} B.M. Boghosian, P.J. Love, P.V. Coveney, I.V. Karlin, S. Succi, 
%and J. Yepez, Phys. Rev. E {\bf 68}, 025103(R) (2003).
%
%\bibitem{ts3} P. Douglas, S. Bergamini, and F. Renzoni, Phys. Rev. Lett.  {\bf 96}, 110601 (2006).
%
%\bibitem{ts4} T.S. Bir\'{o}, and A. Jakov\'{a}c, Phys. Rev. Lett. {\bf 94}, 132302 (2005).
%
%\bibitem{gbiro} G. Biro, G. G. Barnafoldi and T. S. Biro,
%%``Tsallis-thermometer: a QGP indicator for large and small collisional systems,''
%J. Phys. G \textbf{47}, 105002 (2020).

%\bibitem{review} W. Busza, K. Rajagopal and W. van der Schee, arXiv: 1802.04801 [hep-ph].

  \bibitem{Cleymans:2012ya}
   J.~Cleymans and D.~Worku,
  %``Relativistic Thermodynamics: Transverse Momentum Distributions in High-Energy Physics,''
   Eur. Phys. J. A \textbf{48}, 160 (2012).
   
   \bibitem{Kakati:2017xvr}
   S.~K.~Tiwari, S.~Tripathy, R.~Sahoo and N.~Kakati,
   %``Dissipative Properties and Isothermal Compressibility of Hot and Dense Hadron Gas using Non-extensive Statistics,''
   Eur. Phys. J. C \textbf{78}, 938 (2018).
   
   \bibitem{Sjostrand:2006za} 
  T.~Sjostrand, S.~Mrenna and P.~Z.~Skands,
  %``PYTHIA 6.4 Physics and Manual,''
  JHEP {\bf 0605}, 026 (2006).

  \bibitem{PYTHIA8html} PYTHIA8 online manual:(http://home.thep.lu.se/~torbjorn/PYTHIA881html/Welcome.html)

  \bibitem{Ortiz:2013yxa} 
  A.~Ortiz Velasquez, P.~Christiansen, E.~Cuautle Flores, I.~Maldonado Cervantes and G.~Pai\'{c},
  %``Color Reconnection and Flowlike Patterns in $pp$ Collisions,''
  Phys.\ Rev.\ Lett.\  {\bf 111}, 042001 (2013).

  \bibitem{Corke:2010yf} 
  R.~Corke and T.~Sjostrand,
  %``Interleaved Parton Showers and Tuning Prospects,''
  JHEP {\bf 1103}, 032 (2011).

  \bibitem{Aad:2010ac} 
  G.~Aad {\it et al.} [ATLAS Collaboration],
  %``Charged-particle multiplicities in pp interactions measured with the ATLAS detector at the LHC,''
  New J.\ Phys.\  {\bf 13}, 053033 (2011).

  \bibitem{Li:2015jpa} 
  B.~C.~Li, Z.~Zhang, J.~H.~Kang, G.~X.~Zhang and F.~H.~Liu,
  %``Tsallis Statistical Interpretation of Transverse Momentum Spectra in High-Energy pA Collisions,''
  Adv.\ High Energy Phys.\  {\bf 2015}, 741816 (2015).

 \bibitem{Thakur:2016boy} 
  D.~Thakur, S.~Tripathy, P.~Garg, R.~Sahoo and J.~Cleymans,
  %DIF > ``Indication of a Differential Freeze-out in Proton-Proton and Heavy-Ion Collisions at RHIC and LHC energies,''
  Adv.\ High Energy Phys.\ {\bf 2016}, 4149352 (2016).


  \bibitem{Khuntia:2018znt} 
  A.~Khuntia, H.~Sharma, S.~Kumar Tiwari, R.~Sahoo and J.~Cleymans,
  %``Radial flow and differential freeze-out in proton-proton collisions at $\sqrt{s} = 7$ TeV at the LHC,''
  Eur.\ Phys.\ J.\ A {\bf 55}, 3 (2019). 

  \bibitem{Acharya:2018orn} 
  S.~Acharya {\it et al.} [ALICE Collaboration],
  %``Multiplicity dependence of light-flavor hadron production in pp collisions at $\sqrt{s}$ = 7 TeV,''
  Phys.\ Rev.\ C {\bf 99}, 024906 (2019).

% \bibitem{Cuautle:2016ukm} 
% E.~Cuautle, S.~Iga, A.~Ortiz and G.~Paic,
%  %``Color reconnection: a fundamental ingredient of the hadronisation in p-p collisions,''
%  J.\ Phys.\ Conf.\ Ser.\  {\bf 730}, 012009 (2016).

  \bibitem{Sarwar:2015irq} 
  G.~Sarwar, S.~Chatterjee and J.~Alam,
  %``An estimate of the bulk viscosity of the hadronic medium,''
  J.\ Phys.\ G {\bf 44}, 055101 (2017).

\bibitem{Marques:2015mwa}
  L.~Marques, J.~Cleymans and A.~Deppman,
 %``Description of High-Energy $pp$ Collisions Using Tsallis Thermodynamics: Transverse Momentum and Rapidity Distributions,'' 
  Phys. Rev. D \textbf{91}, 054025 (2015).
  
  \bibitem{Cleymans:2012ac}
J.~Cleymans,
%``The Tsallis Distribution at the LHC,''
EPJ Web Conf. \textbf{70}, 00009 (2014).

 \bibitem{Bhattacharyya:2015zka} 
  A.~Bhattacharyya, R.~Ray, S.~Samanta and S.~Sur,
  %``Thermodynamics and fluctuations of conserved charges in a hadron resonance gas model in a finite volume,''
  Phys.\ Rev.\ C {\bf 91}, 041901 (2015).

\bibitem{Aamodt:2011mr}
K.~Aamodt {\it et al.} [ALICE Collaboration],
%``Two-pion Bose-Einstein correlations in central Pb-Pb collisions at $\sqrt{{s}_{NN}} =$ 2.76 TeV,''
Phys. Lett. B \textbf{696}, 328 (2011).

\bibitem{Abelev:2014pja}
B.~B.~Abelev \textit{et al.} [ALICE Collaboration],
%``Freeze-out radii extracted from three-pion cumulants in pp, p\textendash{}Pb and Pb\textendash{}Pb collisions at the LHC,''
Phys. Lett. B \textbf{739}, 139 (2014).

\bibitem{Adam:2016bpr}
J.~Adam \textit{et al.} [ALICE Collaboration],
%``Production of K$^{*}$ (892)$^{0}$ and $\phi $ (1020) in p\textendash{}Pb collisions at $\sqrt{s_{{\text {NN}}}}$ = 5.02 TeV,''
Eur. Phys. J. C \textbf{76}, 245 (2016).

 \bibitem{Saraswat:2017kpg}
  K.~Saraswat, P.~Shukla and V.~Singh,
  %``Transverse momentum spectra of hadrons in high energy pp and heavy ion collisions,''
  J.\ Phys.\ Comm.\  {\bf 2}, 035003 (2018).

  \bibitem{Alkin:2017zob} 
  A.~Alkin,
  %``Phenomenology of charged-particle multiplicity distributions,''
  Ukr.\ J.\ Phys.\  {\bf 62}, 743 (2017).

\end{thebibliography}
\end{document}